\documentclass[fleqn,usenatbib]{mnras}

\usepackage{newtxtext,newtxmath}

\usepackage[T1]{fontenc}
\usepackage{ae,aecompl}
\usepackage[colorinlistoftodos]{todonotes}

\usepackage{graphicx}	
\usepackage{amsmath}

\title[On the Flaring of Thick Discs of Galaxies]{On the Flaring of Thick Discs of Galaxies: Insights from Simulations}

\author[J. García de la Cruz et al.]{
Joaquín García de la Cruz,$^{1}$\thanks{E-mail: j.garciadelacruz@2017.ljmu.ac.uk}
Marie Martig,$^{1}$
Ivan Minchev,$^{2}$
Philip James$^{1}$
\\
$^{1}$Astrophysics Research Institute, Liverpool John Moores University, 146 Brownlow Hill, Liverpool L3 5RF, UK\\
$^{2}$Leibniz-Institut f\"{u}r Astrophysik Potsdam (AIP), An der Sternwarte 16, D-14482, Potsdam, Germany\\
}

\date{Accepted XXX. Received YYY; in original form ZZZ}

\pubyear{2020}

\begin{document}
\label{firstpage}
\pagerange{\pageref{firstpage}--\pageref{lastpage}}
\maketitle

\begin{abstract}
Using simulated galaxies in their cosmological context, we analyse how the flaring of mono-age populations (MAPs) influences the flaring and the age structure of geometrically-defined thick discs. We also explore under which circumstances the geometric thin and thick discs are meaningfully distinct components, or are part of a single continuous structure as in the Milky Way. We find that flat thick discs are created when MAPs barely flare or have low surface density at the radius where they start flaring. When looking at the vertical distribution of MAPs, these galaxies show a continuous thin/thick structure. They also have radial age gradients and tend to have quiescent merger histories. Those characteristics are consistent with what is observed in the Milky Way. Flared thick discs, on the other hand, are created when the MAPs that flare have a high surface density at the radius where they start flaring. The thick discs' scale-heights can either be dominated by multiple MAPs or just a few, depending on the mass and scale-height distribution of the MAPs. In a large fraction of these galaxies, thin and thick discs are clearly distinct structures. Finally, flared thick discs have diverse radial age gradients and merger histories, with galaxies that are more massive or that have undergone massive mergers showing flatter age radial gradients in their thick disc.
\end{abstract}


\section{Introduction}

Thick discs were first discovered in external galaxies by  \citealp{Tsikoudi1979Photometry3115} and \citealp{Burstein1979StructureS0s}, and later on in the Milky Way by \cite{10.1093/mnras/202.4.1025}. Since then, they have been shown to be present in the majority of disc galaxies \citep{Dalcanton2002ADisks,Yoachim2006,Comeron2018}. 

There are many ways to define thick discs. 
The geometrical thick disc refers to a second exponential component in either a disc's vertical light profile  \citep[e.g.,][]{Tsikoudi1979Photometry3115}, or stellar density \citep{Juric2008}. This definition can be used both for external galaxies and the Milky Way. In our Galaxy, many other thin/thick discs distinctions have been proposed based on kinematics \citep{Prochaska2000TheAbundances,Bensby2003,Bensby2005,Reddy2003TheDwarfs}, chemistry  \citep{Fuhrmann1998NearbyHalo, Navarro2011ThroughNeighbourhood, Adibekyan2012}, or age \citep{Haywood2013,Bensby2014ExploringNeighbourhood,Kubryk2015EvolutionDisks}; but for this paper, we will use a geometric definition as it can be applied to all galaxies.

The stellar masses of such geometrically thick discs range from $\mathrm{10^{8.5}}$ to $\mathrm{10^{10.5}}$ $\mathrm{M_{\odot}}$ \citep[e.g.,][]{Comeron2011}, and seem to correlate with their host galaxy's masses, or circular velocities \citep{Comeron2018}. By contrast, the thick-to-thin disc mass ratio, which spans values from 0.3 to 3, is anticorrelated with galaxy mass \citep{Yoachim2006,Comeron2011,Comeron2014,Elmegreen2017ThickFields,Martinez-Lombilla2019PropertiesImaging}.
Scale-heights range from 70 to 410 pc for thin discs, and 330 to 2400 pc for thick discs \citep{Comeron2011}. 
In the Milky Way, \citealp{Juric2008} find the scale-heights for the thin and thick discs to be $\sim$ 300 and $\sim$ 900 pc respectively, but more recent estimations of the thick disc scale height are around $\sim$ 500-700 pc (\citealp{Robin2014,Mateu2018TheStars}, see also \citealp{Bland-Hawthorn2016TheProperties} and references therein). 
Among other thick disc properties, one this work focuses on is the global shape of the disc, i.e. the possible variations of scale-height with radius.
Studies on large samples of galaxies  \citep[e.g.,][]{Yoachim2006,Comeron2011,Comeron2012,Comeron2014,Comeron2018} concluded that the vast majority of thick discs in external galaxies have a constant scale-height with radius, even though other works had suggested that flat thick discs should belong to late-type galaxies only \citep{DeGrijs1997TheDistance,Bizyaev2014THEDISKS}. 
The studies which found only flat thick discs, however, either show one averaged scale-height for the whole thick disc \citep{Yoachim2006}, or
describe the disc with only a few radial bins
\citep{Comeron2011,Comeron2012,Comeron2014,Comeron2018}, making it difficult to observe variations in the scale-heights with galactic radius. 
This does not mean that thick discs are never flat: this has been clearly shown in some cases, for instance by \citealp{Ibata2009} and \citealp{Streich2016ExtragalacticGalaxies}.
However, other works have also found thick discs with a significant amount of flaring \citep{Narayan2002OriginDisk,Kasparova2016,Sarkar2019Flaring7321,Kasparova2020An7572}, and \citealp{Rich2019TheDiameters} found boxy shaped isophotes in the outer disc of some edge-on galaxies, which indicate strong disc flaring \citep{10.1093/mnras/staa678}.

In the Milky Way, although flaring in the inner disc ($\sim$11 kpc) is highly disfavoured \citep{Mateu2018TheStars}, the existence of thick disc flaring in the outskirts \citep{Robin2014ConstrainingWay,Mateu2018TheStars,Lopez-Corredoira2018,10.1093/mnras/sty2604}, and the amplitude of the effect \citep{Reyle2009TheCounts,Polido2013APLANE,Kalberla2014DOESDATA,Lopez-Corredoira2014FlareData,Amores2017EvolutionShape} are still a matter of debate.
In any case, there is increasing evidence that thick discs might be more diverse than previously thought in terms of global shape and flaring. 

Similarly, the age structures created by the stellar populations in thick discs are also showing more diversity in recent studies than previously thought. 
Some works using broadband photometry \citep[e.g.,][]{Dalcanton2002ADisks,Seth2005APopulation,Mould2005RedGalaxies} and spectroscopy \citep{Yoachim2008LickGalaxies,Comeron2015,Comeron2016} helped establish the classical view of thick discs as red, old, and metal poor components. 
However, on the one hand, colour cannot account for the age-metallicity degeneracy, especially for old ages (>6 Gyr).
On the other hand, spectroscopy is limited by the low surface brightness of the outer regions of the thick disc (see a more detailed discussion in section 4.2 in \citealp{Martig2016AWAY}).
In spite of these limitations, more recent works suggest that thick discs are diverse and have complex age structures.
\citealp{Kasparova2016} analysed spectroscopically three edge-on galaxies, and found that while one has a very old thick disc ($\sim$10 Gyr), the other two have intermediate age ($\sim$5 Gyr) stars in their thick discs.
In the Fornax cluster, \citealp{Pinna2019TheAccretion,Pinna2019TheEnvironment} found that in FCC 170 (NGC 1381), FCC153 and FCC 177, the thick discs are very old, but host complex populations.
In the Milky Way, the geometrically thick disc is made of old stars in the solar neighbourhood and the inner disc, but shows a strong radial age gradient: between galactocentric radii of 6 and 12 kpc, the mean age of red clump stars in the thick disc drops from $\sim 9$ to $\sim 5$ Gyr (\citealp{Martig2016AWAY}, see also \citealp{Ness2016SPECTROSCOPICGIANTS,Xiang2017TheSurveys,Feuillet2019SpatialRelation}). Such a radial age gradient has not been found yet in any other galaxy, either because the few galaxies that have been observed do not have the same structure as the Milky Way, or because the observations were not able to probe the very outer regions of the thick discs. In any case, a picture is emerging where thick discs can host complex stellar populations, have different age structures, and also possibly different relationships with the thin discs.

The age structure of thick discs is also connected to the important question of their relationship to thin discs: are thin and thick discs clearly distinct components, and is the separation meaningful?
\citealp{Streich2016ExtragalacticGalaxies} do not find a separate thick disc structure for three external galaxies, 
while MUSE observations by \citealp{Guerou2016ExploringMUSE} or \citealp{Pinna2019TheAccretion} found thick discs that seem clearly distinct from the thin disc. The situation is more complex in the Milky Way. It has been suggested for a long time that the thin and thick discs could just be part of the same continuous structure \citep{Nemec1991MixtureComponents,Nemec1993MixtureModels,Norris1999TheStatus}. \citealp{Bovy2012a} have then shown that the transition from thin to thick disc corresponds to a continuum of stellar populations, and that the Milky Way's thick disc is not a distinct component. \citealp{Rix2013TheDisk} further argue that while the vertical density profile can be very well fitted by a sum of two exponentials, these do not correspond to a meaningful physical decomposition (see also \citealp{Park2020ExploringSimulations} for a discussion of this issue using numerical simulations). Whether the Milky Way is a unique case or not is still unknown, but what is most likely is that there exists a variety of disc structures.
Ultimately, the age structure of thick discs and the thin/thick disc dichotomy must be linked to the formation process of galactic discs.

Thick discs have been proposed to be a natural consequence of $\mathrm{\Lambda}$CDM \citep{Read2008d}.
Yet, their exact origin is still a matter of debate.
One mechanism for thick disc formation is radial migration, proposed by \citealp{Schonrich2009,Loebman2011TheMigration}. However, later on, works like \citealp{Minchev2012a,Martig2014}, and \citealp{Vera-Ciro2014} found that radial migration does not contribute much to disc thickening and, on the contrary, it suppresses flaring when external perturbations are included \citep{Minchev2014,Grand2016VerticalContext}. 
Other proposed mechanisms are thick discs originating through stars born out of accreted gas from satellites \citep{Brook2004}, accreted stars from galaxy mergers \citep{Abadi2003}, or from disc-crossing satellites heating up the thin disc \citep{Quinn1993,Read2008d,Villalobos2008SimulationsDiscs}. 
\citealp{Bournaud2009} proposed that flat thick discs could form through a clumpy and turbulent phase in the evolution of galaxies at early times.
Observations of flat thick discs favoured this scenario, but it remained unknown why mergers (an inevitable consequence of $\Lambda$CDM) did not produce disc flaring as seen in the simulations \citep{Kazantzidis2008ColdAccretion,Kazantzidis2009ColdAccretion,Qu2011CharacteristicsLengths,Moetazedian2016ImpactDisc}. 

\citealp{Minchev2015ONDISKS} proposed a solution for this conundrum, showing that a flat thick disc can be recovered by superposing several stellar mono-age populations, MAPs \footnote{We note that ‘MAPs’ has been used extensively in the past to mean mono-abundance populations (e.g., \citealp{Bovy2012,Rix2013TheDisk}), but those can be very different from mono-age population, and should not be confused with them, as discussed in detail by \citealp{Minchev2016THEDISK}.}, with different levels of flaring and radial density profiles. 
The idea of many MAPs with different levels of flaring that nevertheless can produce a flat thick disc when combined, shows that thick discs are indeed complex and probably do not form in a single event.
For their work, \citealp{Minchev2015ONDISKS} used two simulated galaxies, one from the sample we use in this work \citep{Martig2012}, the other from the Aquarius Project haloes \citep{Springel2008,Scannapieco2009}.
This same effect has been found in other sets of simulations as well, like LATTE \citep{Ma2017}, AURIGA \citep{Grand2014}, NIHAO-UHD \citep{Buck2020OnMigration}, and VINTERGATAN \citep{Agertz2020VINTERGATANGalaxy}, and in observations of the Milky Way \citep{Xiang2018StellarLAMOST}. The superposition of MAPs with different levels of flaring is predicted to create a radial age gradient in the thick disc \citep{Minchev2015ONDISKS}, which was observed in the Milky Way \citep{Martig2016AWAY}.

In this paper, we expand on the work of \citealp{Minchev2015ONDISKS} by studying the structure of discs in the full sample of simulated galaxies presented in \citealp{Martig2012}.
We first present some general properties of the simulated galaxies and their discs, but the main focus of the paper is to explore how much MAPs flare, and what kind of thick disc they create.
We pay special attention to three aspects: a) how flaring MAPs can create either a flat or flared thick disc, b) whether those thick discs define a separate structure from the thin disc or not (in terms of their stellar populations), c) and the different kinds of age gradients MAPs create. Finally, we look into the merger histories of our galaxies, and show the connection between mergers and different thick disc properties.

\begin{figure*}
\centering
\includegraphics[width=\textwidth]{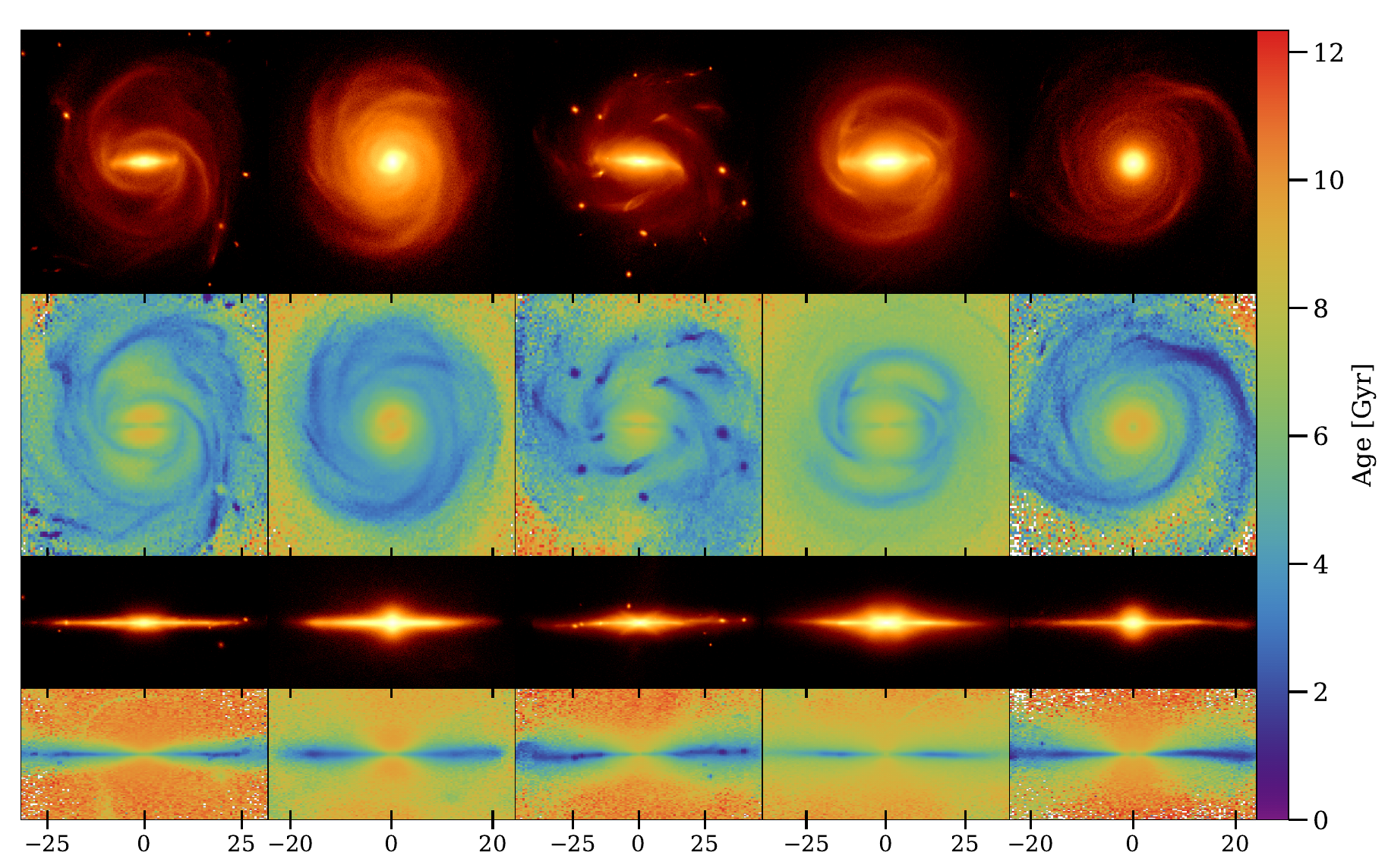}
\caption{Density and age maps seen face-on and edge-on of five galaxies from our sample. From left to right: g92, g47, g36, g39, and g102. We choose these same galaxies as representatives of the scenarios discussed in Sec. \ref{sec:monoagepopulations} and \ref{sec:agegradients}.}
\centering
\label{fig:galsample}
\end{figure*}

\section{Methods}
\label{sec:methods}

\subsection{Simulations}
\label{sec:simulation}
We use a sample of zoom-in numerical re-simulations of galaxies in their cosmological context described in more detail in \citealp{Martig2012}. 
The simulation technique consists of two different stages (see \citealp{Martig2009}). 
In the first stage, a dark-matter-only $\Lambda$CDM cosmological simulation with $512^3$ dark matter particles is run using the adaptive mesh refinement code \texttt{RAMSES} \citep{Teyssier2002}. 
We identify dark matter haloes with a final mass between 2.7$\times 10^{11}$ and 2$\times 10^{12}$ $\mathrm{M_{\odot}}$ that live in isolated environments. 
Then, merger histories and diffuse dark matter accretion are recorded for those halos from $z=5$ to $z=0$, storing the time, mass, velocity, and spin of the satellites. 

In the second stage, a new simulation follows the growth of a seed galaxy which evolves from $\mathrm{z=5}$ to $\mathrm{z=0}$ using the merger and accretion histories obtained in the first simulation. Each incoming halo is replaced by a galaxy containing stars, gas, and dark matter. This new simulation uses the Particle-Mesh code described by \citealp{Bournaud2002,Bournaud2003} with a sticky-particle algorithm for modelling gas dynamics. 
The total size of the box is 800 $\times$ 800 $\times$ 800 $\mathrm{kpc}$, and the spatial resolution is 150 pc. The mass resolution is 1.5$\times 10^{4}$ $\mathrm{M_{\odot}}$ for gas particles and stellar particles formed during the simulation, 7.5$\times 10^{4}$ $\mathrm{M_{\odot}}$ for stellar particles in the initial seed galaxies at $z=5$,
 and 3$\times 10^{5}$  $\mathrm{M_{\odot}}$ for dark matter particles. 
Star formation obeys a Schmidt-Kennicutt law \citep{KennicuttJr.1998}, and energy feedback from supernovae is included. 
A mass loss scheme is implemented following that of \citealp{Jungwiert2001} and used in \citealp{Martig2010}. 
A full detailed explanation of the simulation technique is in Appendix A of \citealp{Martig2009}.

\subsection{The sample}
\label{sec:sample}
The original sample consists of 33 simulated galaxies presented in \citealp{Martig2012} with stellar masses between $10^{10}$ and $2 \times 10^{11}$ $\mathrm{M_{\odot}}$, and a large diversity of morphologies and formation histories.
Some of the galaxies in the sample have structural features
that would vastly complicate our analysis: these are galaxies either undergoing a massive merger at z=0, or having a polar ring, or presenting a dramatically warped and distorted disc. 
For that reason, in this work we exclude 6 galaxies from the original sample, leaving us with 27 galaxies (see some examples in Figure \ref{fig:galsample}). 

\subsection{Scale-heights}
\label{sec:scaleheight_fits}

We first compute a global estimate of the overall disc thickness: we define $h_\mathrm{scale}$ as the standard deviation of the vertical position of stars located at half the optical radius of the galaxy, $R_{25}$.
To then compute the scale-heights of the global thin and thick disc, we bin the disc stellar particles in cylindrical shells with a width of 2 kpc and a height of 3 $h_\mathrm{scale}$. 
At each radius, we compute the vertical number density of particles using 20 bins, and fit the profile using a combination of two $\mathrm{sech^2}$ functions:

\begin{equation}
    N(z)=N_{0} \left((1-\alpha)\textrm{sech}^2\left(\frac{z}{h_\mathrm{thin}}\right)+\alpha  \textrm{sech}^2\left(\frac{z}{h_\mathrm{Thick}}\right)\right)
\end{equation}

where $N_{\mathrm{0}}$ is the stellar number density at the mid-plane, $h_{\mathrm{thin}}$ and $h_{\mathrm{Thick}}$ correspond to the scale-heights of the thin and thick disc respectively, and $\mathrm{\alpha}$ is the number density fraction of the thick disc over the global disc. We find that a single $\mathrm{sech^2}$ could not describe well the vertical distribution of stars, especially for the thickest discs.
To determine the vertical stellar distribution of stars, following \cite{Bennett2019VerticalDR2}, we use a Poisson distribution for the likelihood,  which we write as follows:

\begin{equation}
    \textrm{ln}\mathcal{L}(N_{c}|N_{p})=\sum-N_{p}+N_{c}\cdot \textrm{ln}(N_p) - \textrm{ln}(N_{c}!) 
\end{equation}

where ${N_\mathrm{c}}$ is the number of stellar particles counted in the bin, $N_{\mathrm{p}}$ is the number of stellar particles predicted by the model, and $N_{\mathrm{c}}!$ is independent of the models and thus ignored for the computation of the likelihood.

\begin{figure}
\centering
\includegraphics[width=\columnwidth]{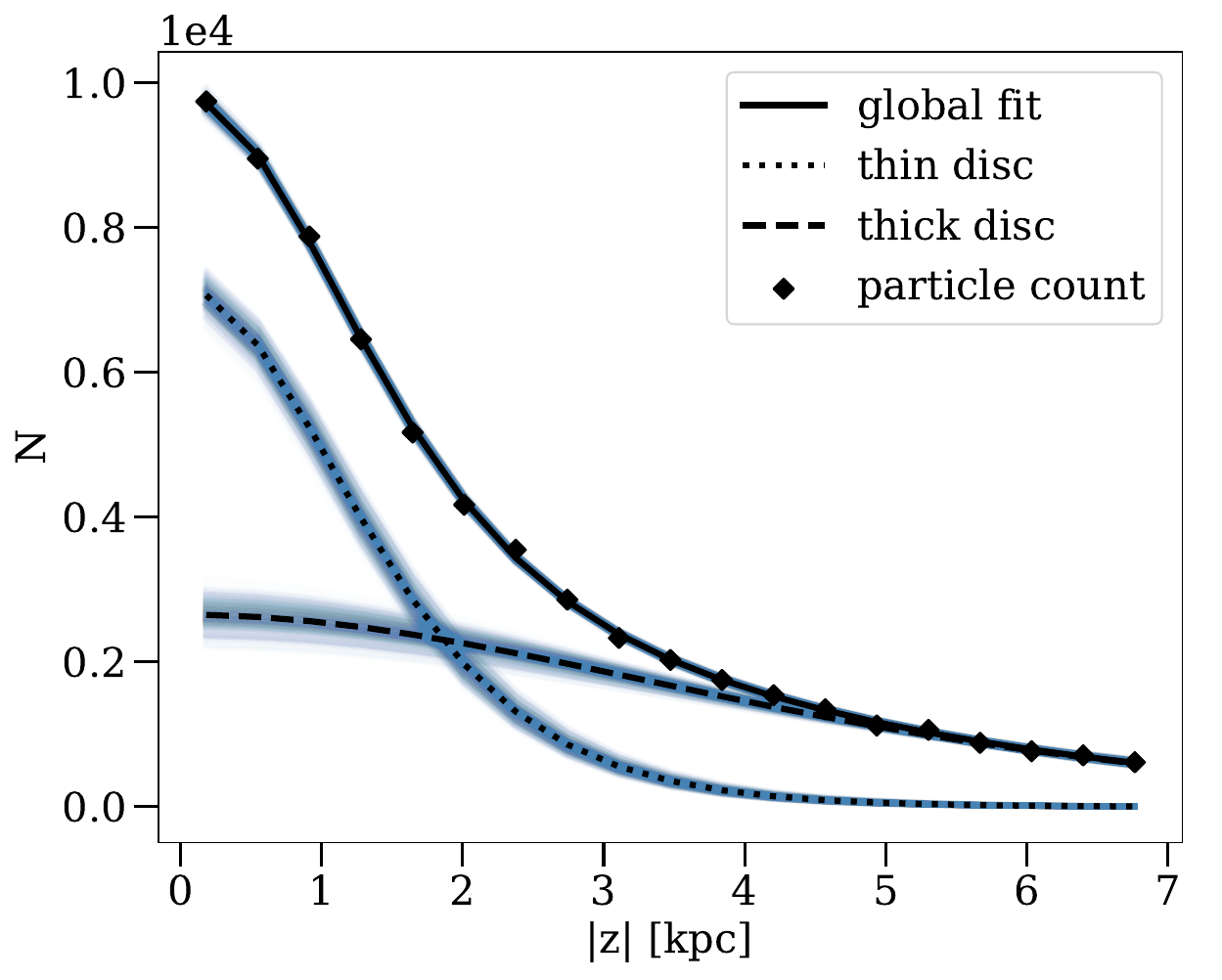}
\includegraphics[width=\columnwidth]{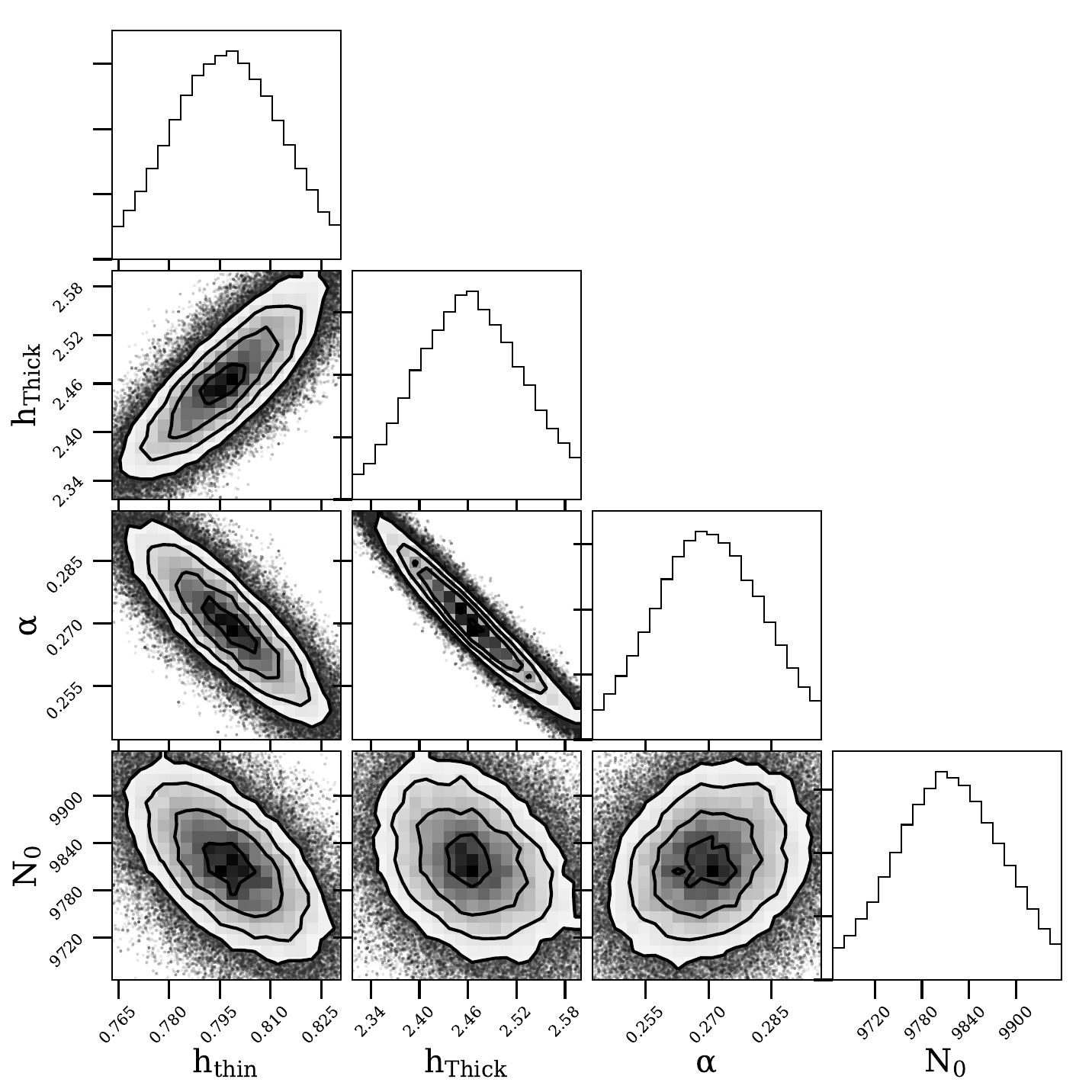}
\caption{\textit{Top}: Vertical distribution of number density of stellar particles in black diamonds and the MCMC fit in solid black line. 
The dotted and dashed line represent the thin and thick disc contribution to the total particle count. The blue range indicates the 16$^\mathrm{th}$ to 84$^\mathrm{th}$ percentiles ranges of the posterior distribution.
\textit{Bottom}: corner plot of the parameter values for the MCMC fit to the vertical number density profile presented in the upper panel.} 
\centering
\label{fig:MCMC}
\end{figure}

The fits for the scale-height were obtained using Markov Chain Monte Carlo (MCMC) with the \texttt{python} package \texttt{emcee} \citep{Foreman-Mackey2013EmceeHammer}, with 200 walkers and 5000 steps. The walkers start from random positions around the best fit value obtained using the  \texttt{ScyPy} routine \texttt{curve\_fit} \citep{Virtanen2019SciPyPython}.
For the priors, we let $h_{\mathrm{thin}}$ and $h_{\mathrm{Thick}}$ take any value from 0 to 15 kpc, $\alpha$ from 0 to 1, and we set $N_{0}$ to be positive. 
The final values we report for each parameter are the median and the 16$^\mathrm{th}$ to 84$^\mathrm{th}$ percentiles range of the posterior distribution. An example of the MCMC fits can be seen in Fig. \ref{fig:MCMC}.

In addition to fits to the thin and thick discs, we also determine the scale-heights of MAPs, for which we split the stellar particles into 0.5 Gyr age bins ranging from 0 to 13 Gyr old, and the same spatial bins as for the fits of the global thin and thick discs. 
\citealp{Martig2014} showed that a single exponential provides a good fit to the vertical density profile of MAPs, as was also found by \citealp{Minchev2015ONDISKS}. 
Nevertheless, since we used a squared hyperbolic secant to fit the thin and thick discs, we used this same function to fit the MAPs in order for both MAP and thin/thick global scale-heights to be comparable.  
The fits for the MAPs' scale-heights are done in each radial bin, from the galactic centre to $R_{25}$ until either there are fewer than 10 stellar particles inside the bin or the scale-heights are greater than 10 kpc.

\subsection{Radial extent of thick discs}
\label{sec:radial_extent}

Although the scale-height fits are initially done all the way from the centre to $R_{25}$, we are interested in the disc only. 
The fits in the very inner parts of the galaxy most likely represent the vertical density distribution of other components ---e.g. the bulge and the bar if the galaxy has one--- which are not within the scope of this paper. 

\begin{figure}
\centering
\includegraphics[width=\columnwidth]{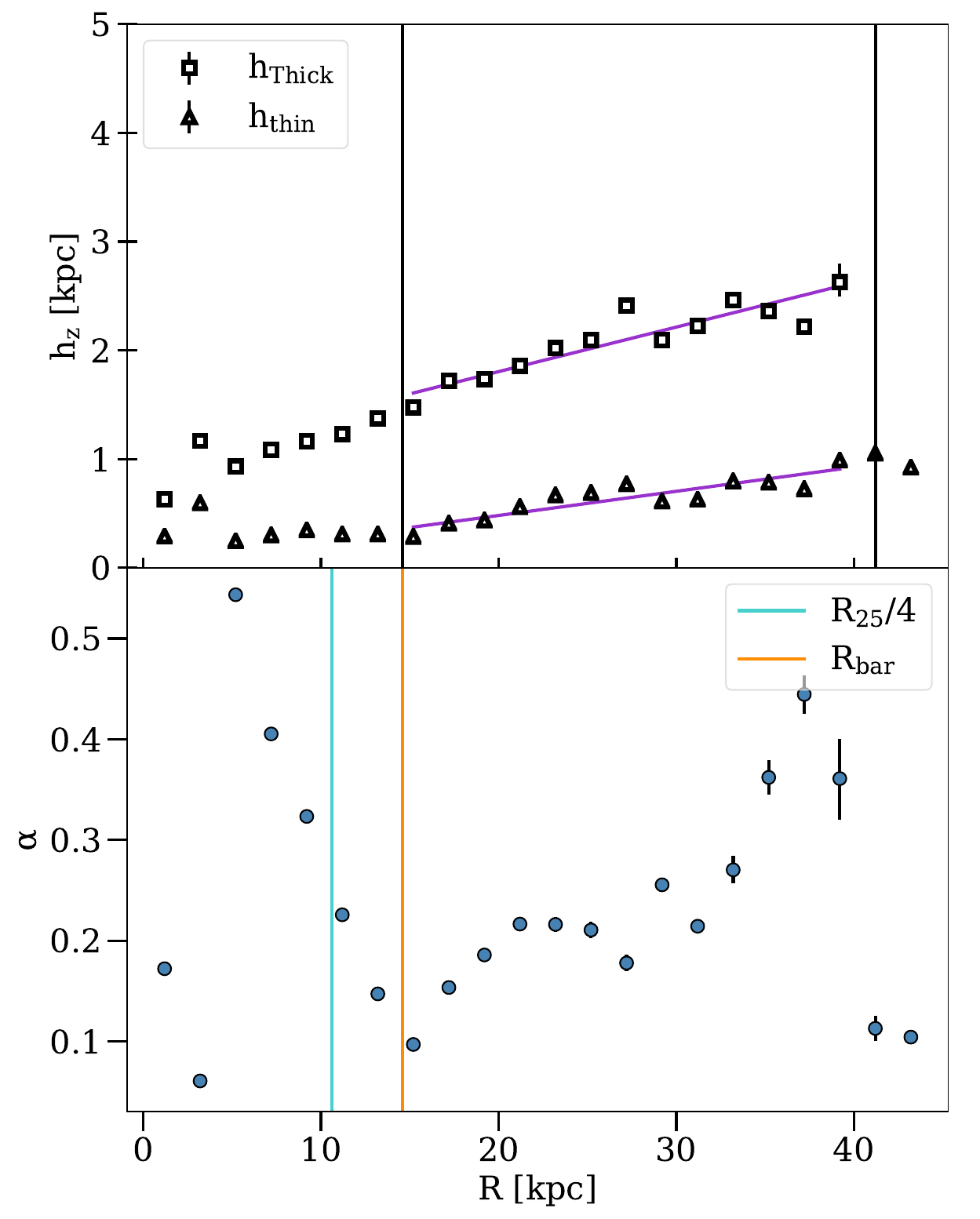}
\caption{\textit{Top}: radial profile of the thin and thick discs' scale-heights for galaxy g36. The two vertical black lines represent the inner \textbf{$R_{\mathrm{inner}}$} and outer \textbf{$R_{\mathrm{outer}}$} boundaries of the disc, whereas the purple lines represent the linear fits to the scale-heights of the two components of the disc. \textit{Bottom}: Radial distribution of the thick disc fraction. 
We see a change in the  behaviour of $\alpha$ in the inner disc, corresponding to the end of the bar (we use $R_{\mathrm{bar}}$ to define the inner boundary of the thick disc), and we also see a drop at the outer edge of the thick disc.} 

\centering
\label{fig:radexet}
\end{figure}

\begin{figure}
\centering
\includegraphics[width=\columnwidth]{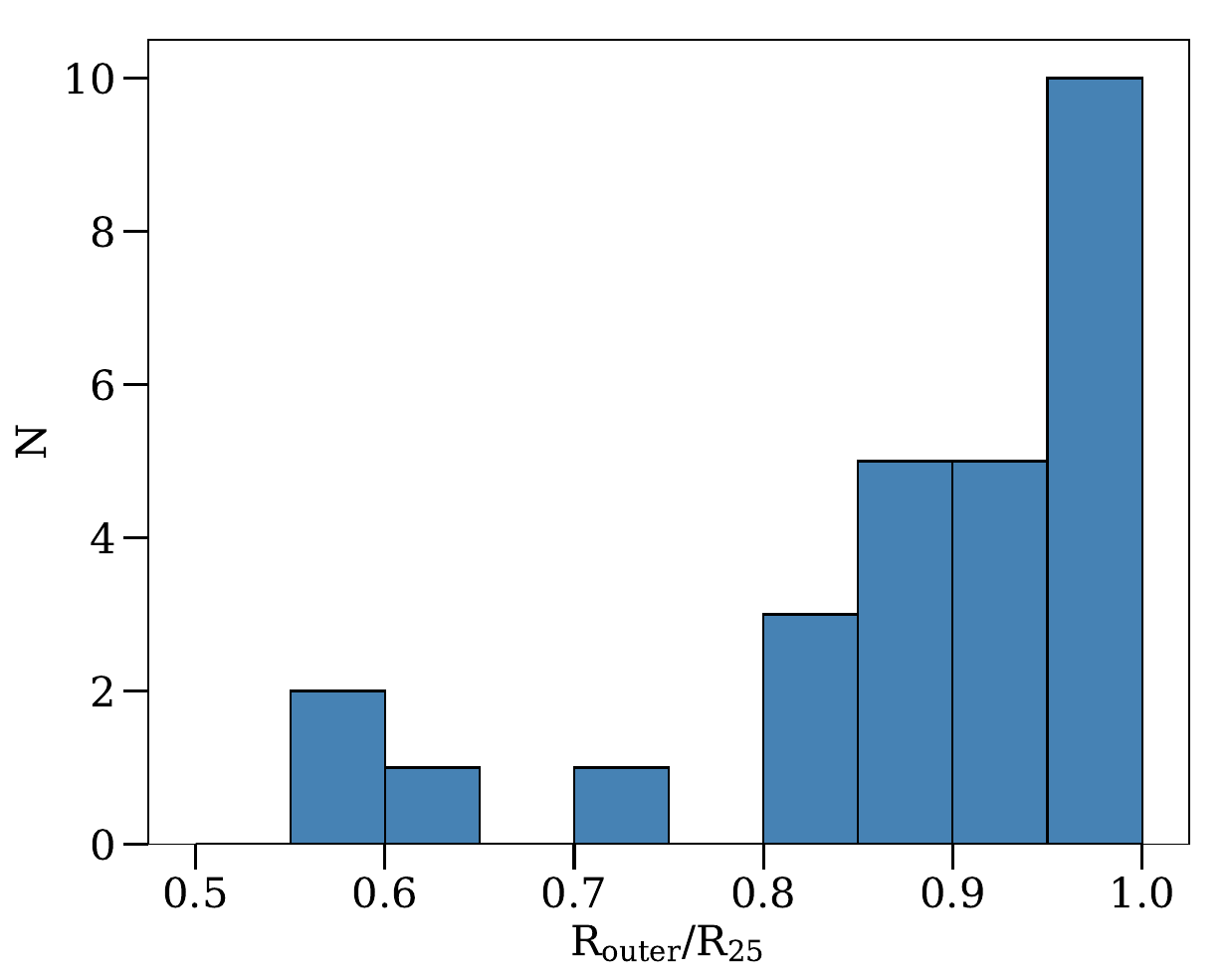}
\caption{Histogram representing the radial extent of the thick disc over $R_{25}$ after applying our criteria. For 15 galaxies the thick disc extends beyond 90\% of $R_{25}$.}
\centering
\label{fig:radexet_hist}
\end{figure}

In order to separate the central part of the galaxy from the disc, we study the density maps of the galaxies and see that the bulges end at around a quarter of $R_{25}$.
We also noticed that the behaviour of the thick disc fraction, $\alpha$, changes at around the same radius.
This can be explained if indeed the $\mathrm{sech^2}$ are describing different galaxy components inside and outside of this radius. 
For some galaxies, the bar goes beyond $R_{25}/4$. 
In these cases, the changing in trend of $\alpha$ is not at $R_{25}/4$, but at the bar's end. 
Therefore, we consider the inner boundary of the disc, $R_{\mathrm{inner}}$ hereafter, to be at a quarter of $R_{25}$ for most galaxies, or at the bar's end if it extends beyond $R_{25}/4$.

For some galaxies, the MCMC fits in the outer part of the disc show either very large errors or values significantly higher than for the rest of the disc. 
When we look at the behaviour of $\alpha$ in the same region, it shows very low values, indicating that the thick disc component is marginal in terms of mass.
The reasons behind this change of behaviour or poor fits at large radii can be multiple: a break in the thick disc, the vertical density follows a different distribution than a $\mathrm{sech^2}$, a warp, etc. 
We define the outer limit of the thick disc, hereafter $R_{\mathrm{outer}}$, to be this radius at which the fit starts to perform poorly.
An example of how these criteria were applied and their result is shown in Fig. \ref{fig:radexet}.
In Fig. \ref{fig:radexet_hist} we present the distribution of $R_{\mathrm{outer}}$/$R_{25}$. In most cases, we find that thick discs have nearly the same extent as the overall disc: for 15 out of 27 galaxies, $R_{\mathrm{outer}}$ is greater than 0.9 $R_{25}$.

The total mass of each disc component is the sum of the corresponding $\mathrm{sech^2}$ at every radial bin from $R_{\mathrm{inner}}$ to $R_{25}$ for the thin disc, and from $R_{\mathrm{inner}}$ to $R_{\mathrm{outer}}$ for the thick disc. The total mass of the disc, $M_{\mathrm{disc}}$, is the sum of the mass of the thin and thick discs. Hereafter, we denote the latter $M_{\mathrm{Thick}}$.

\subsection{Fitting the disc flaring}

To quantify the level of flaring for the thin and thick discs, we study how their scale-heights change with radius.
We find that when taking into account the radial extent criteria explained in the previous subsection, the radial profiles of both scale-heights are well described by a single line (see Figure \ref{fig:radexet}), and the slope of that line can be used to quantify the level of flaring.
Therefore, we use the \texttt{least\_squares} algorithm from \texttt{ScyPy} \citep{Virtanen2019SciPyPython} to perform a linear fit to the radial profiles of both the thin and the thick discs' scale-heights. We use the Cauchy loss function in order to minimise the effect of outliers. The respective slopes $\nabla_{\mathrm{thin}}$ and $\nabla_{\mathrm{Thick}}$, will be hereafter the thin and thick disc scale-height gradients respectively.  

\section{General properties}
\label{sec:generalproperties}
\begin{figure*}
\centering
\includegraphics[scale=0.36]{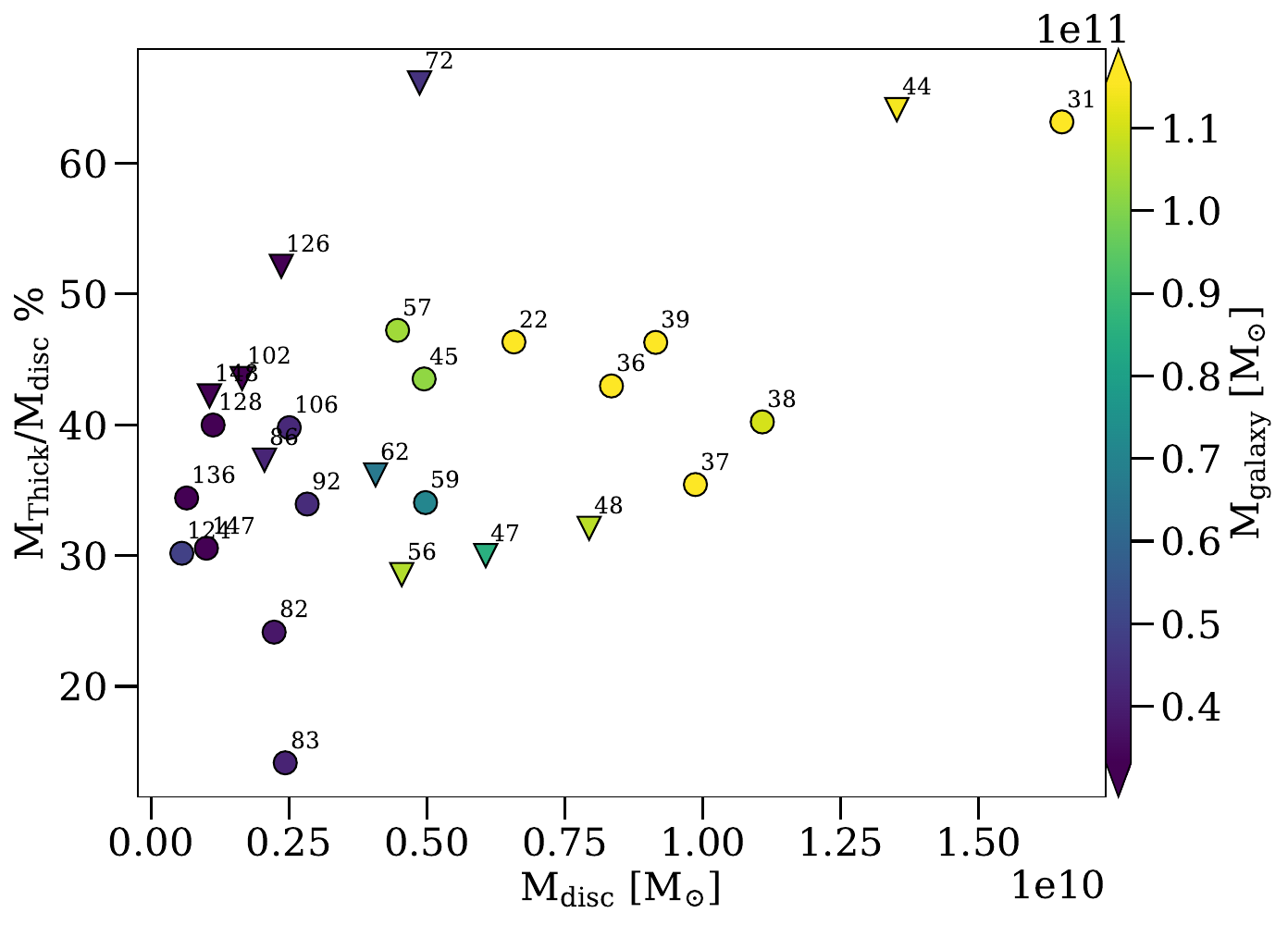}
\includegraphics[scale=0.36]{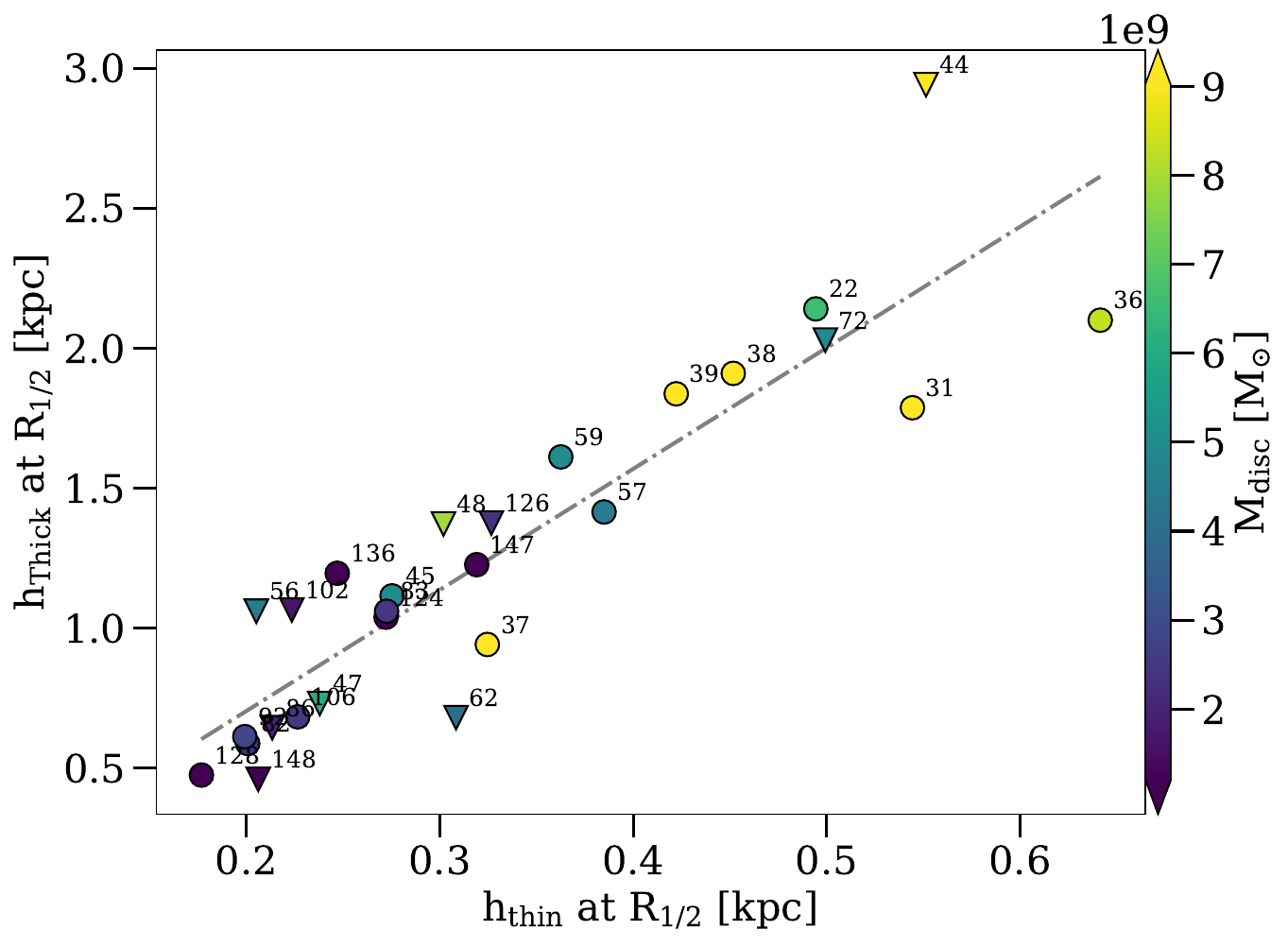}
\includegraphics[scale=0.36]{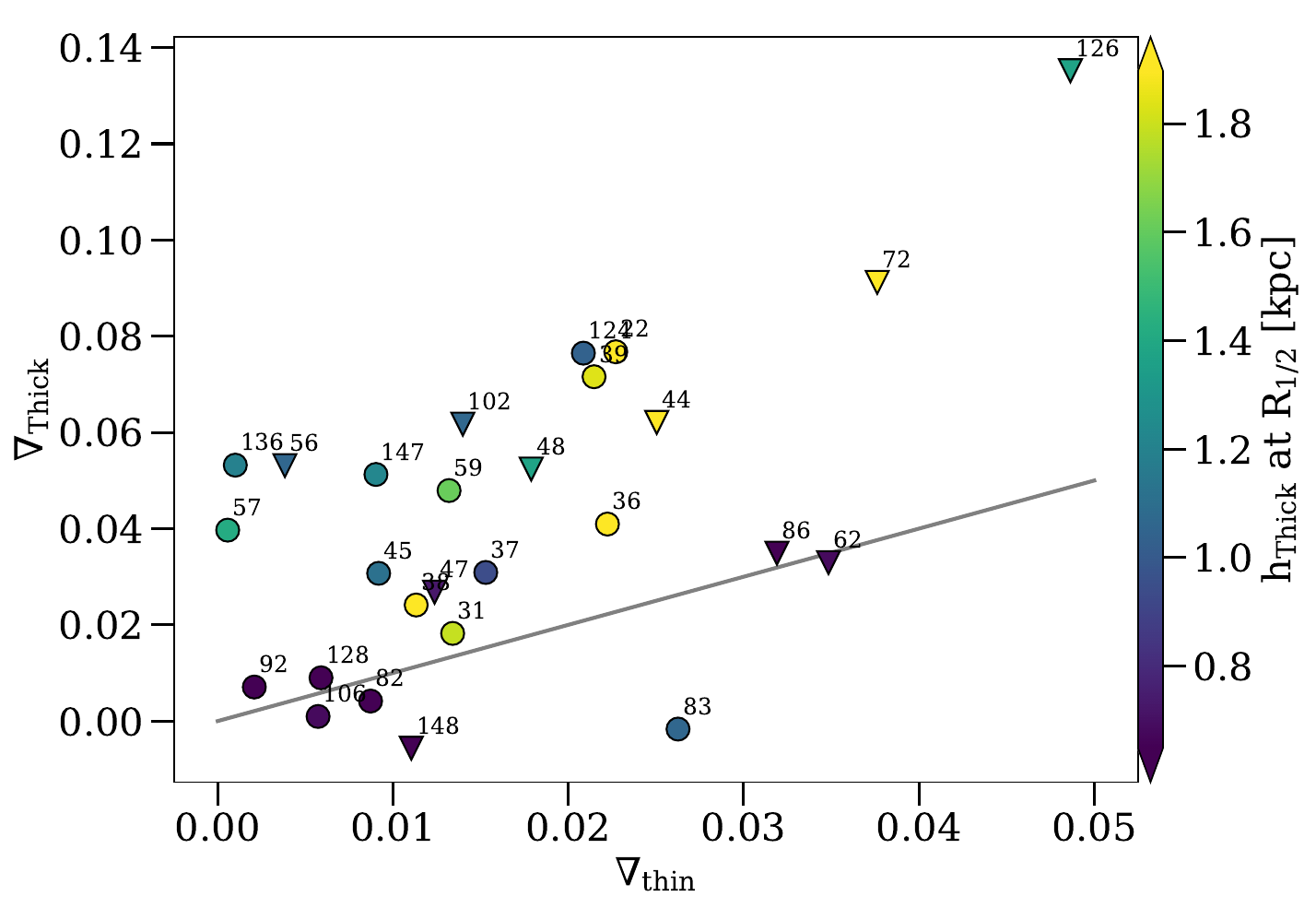}
\includegraphics[scale=0.36]{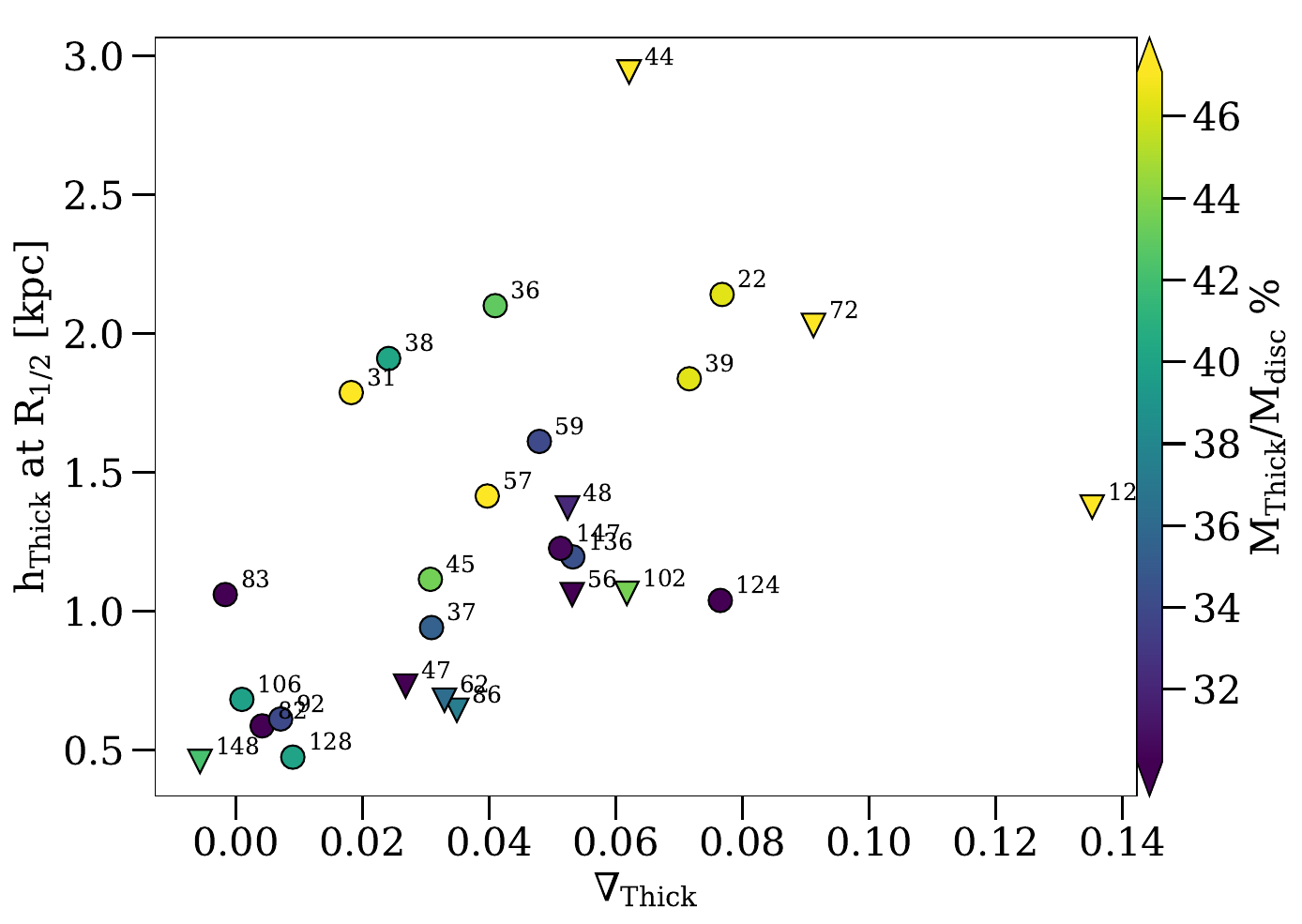}
\caption{\textit{Top left:} thick disc mass ratio against total disc mass, colour-coded by the stellar mass of the galaxy. \textit{Top right:} scale-height of the thick disc at the middle of the disc against scale-height of the thin disc at the middle of the disc, colour-coded by the mass of the disc. The gray line represents a linear fit to the data. \textit{Bottom left:} thick disc scale-height gradient against thin disc scale-height gradient, colour-coded by the scale-height of the thick disc at the middle of the disc. The gray line is a 1:1 line to show that thick discs are more flared than thin discs in most cases. \textit{Bottom right:} scale-height of the thick disc at the middle of the disc against thick disc scale-height gradient, colour-coded by the thick disc mass ratio.
The number on the top right of each data point is the galaxy number/label.  Triangles indicate that the extent of the thick disc is above 0.9 of $R_{25}$, and circles otherwise. The arrow-shaped ends of the colourbar indicate that values range from the 16th to the 84th percentile of the variable.}
\centering
\label{fig:generalproperties}
\end{figure*}

In this section, we study the global properties of the discs, i.e. their mass, thickness, and degree of flaring.
The upper left panel of Fig. \ref{fig:generalproperties} shows the thick disc stellar mass fraction $M_\mathrm{{Thick}}/{M_\mathrm{disc}}$ as a function of the total disc stellar mass ${M_{\mathrm{disc}}}$, colour-coded by the total stellar mass of the galaxy ${M_{\mathrm{galaxy}}}$ (the symbols represent the radial extent of the thick disc, as will be discussed later). 
The stellar mass ratio of the thick disc varies from cases like g82 and g83 where the mass fraction is below 25\%, to cases like g72, g44, or g31, whose mass fraction is above 60\% of the total disc stellar mass. 
The disc stellar mass correlates strongly with the stellar mass of the galaxy, but it does not correlate strongly with the thick disc stellar mass ratio.  
It seems that low mass discs can have a wide range of thick disc stellar mass ratios, and that high mass discs tend to have slightly higher thick disc ratios on average, but the trend is quite weak and we do not have enough simulated galaxies to confirm this effect.

In addition to disc stellar mass, we also want to quantify the thickness of both the thin and thick discs, which can vary significantly with radius in flared discs. 
In this paper we define the thicknesses as the scale-height $h_{\mathrm{Thick}}$ value at half the extent of the disc, which we denote as $R_{1/2}$. In the upper right panel of Fig. \ref{fig:generalproperties}, we compare the thicknesses of the thin and thick discs, colour-coded by the mass of the disc $M_{\mathrm{disc}}$. The respective thicknesses of the thin and thick discs are tightly correlated: thick discs are approximately 4 times thicker than thin discs. We also find that low mass discs tend to be relatively thinner, with some exceptions (which shows that mass is not the only factor controlling disc structure).

In the bottom left panel of Fig. \ref{fig:generalproperties}, we plot the slope of the thick disc $\nabla_{\mathrm{Thick}}$ against the slope of the thin disc $\nabla_{\mathrm{thin}}$, colour-coded by the thickness of the disc. Overall, thick discs tend to be more flared than thin discs and the values of both slopes mildly correlate with each other.
Also, in general trends, galaxies with a flatter thick disc tend to be thinner, and galaxies with a thinner thick disc have similar slopes for thin and thick discs.

Finally, the thick disc general properties studied in this section are summarised in the bottom right panel of Fig. \ref{fig:generalproperties}. The flaring of the thick disc and its thickness are correlated, and are also correlated with the thick disc mass ratio. 
A final aspect that we investigate is whether the radial extent of thick discs is correlated with any of the properties we have mentioned so far. To this end, in all panels of Fig. \ref{fig:generalproperties}, we represent with triangles galaxies where the thick disc extends out to more than 90\% of $R_{25}$, and with circles galaxies with a less extended thick disc. We do not find any difference between the two populations, and for the rest of the paper we do not separate galaxies into categories based on the extents of their thick discs.

Our sample of galaxies shows a large range of mass, thickness, and flaring of thin and thick discs. From galaxies whose thick discs are quite flat, to galaxies with very flared thick discs, we observe a continuum with no clear gaps or separated categories. Hereafter, although we will refer to flat and flared thick discs, the reader should keep in mind that there are intermediate cases and no clear threshold separates one category from the other. 

\section{Mono-Age Populations and thick disc connection} 
\label{sec:monoagepopulations}

\begin{figure*}
\centering
\includegraphics[scale=0.35]{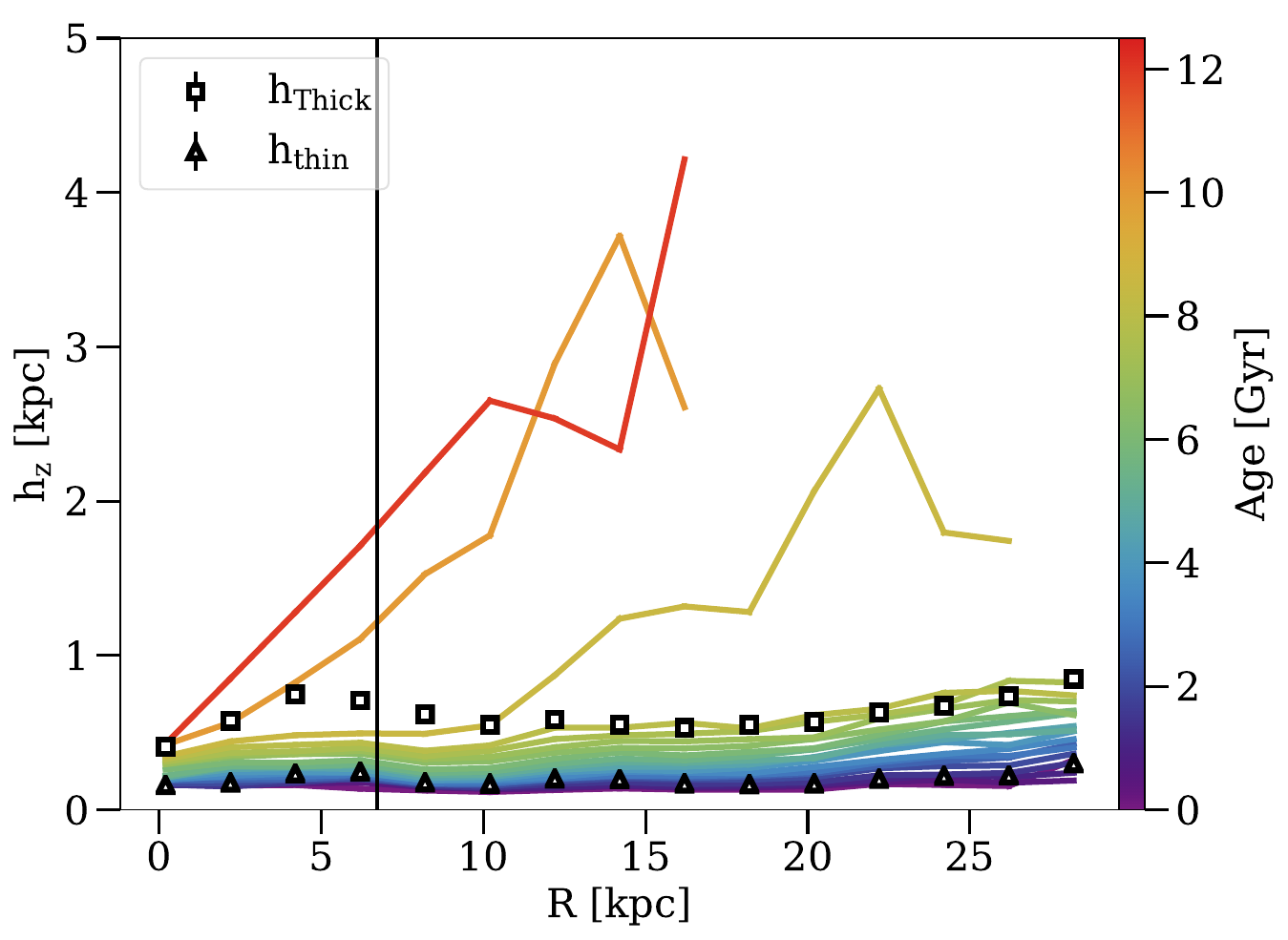}
\includegraphics[scale=0.35]{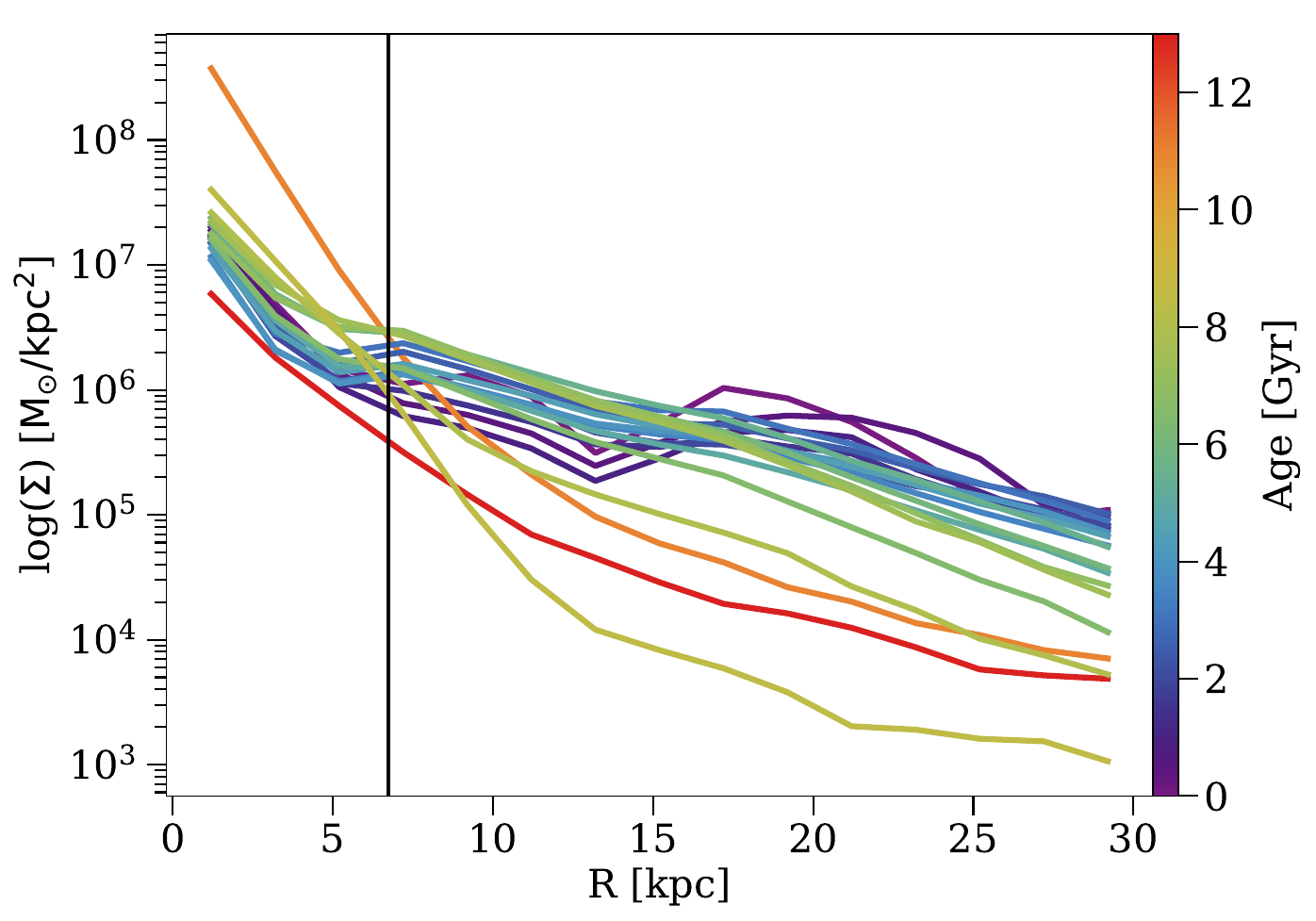}
\includegraphics[scale=0.35]{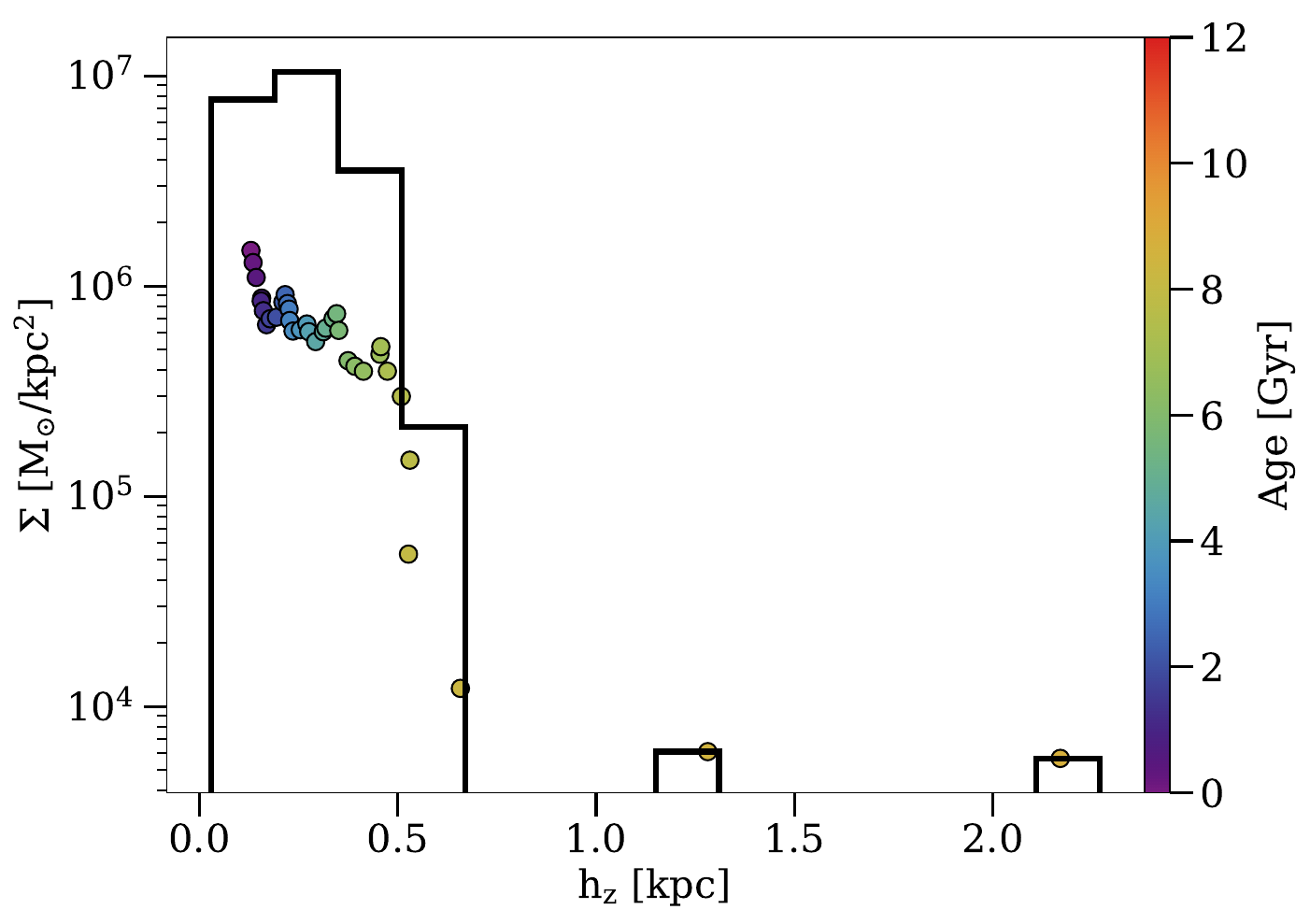}
\includegraphics[scale=0.35]{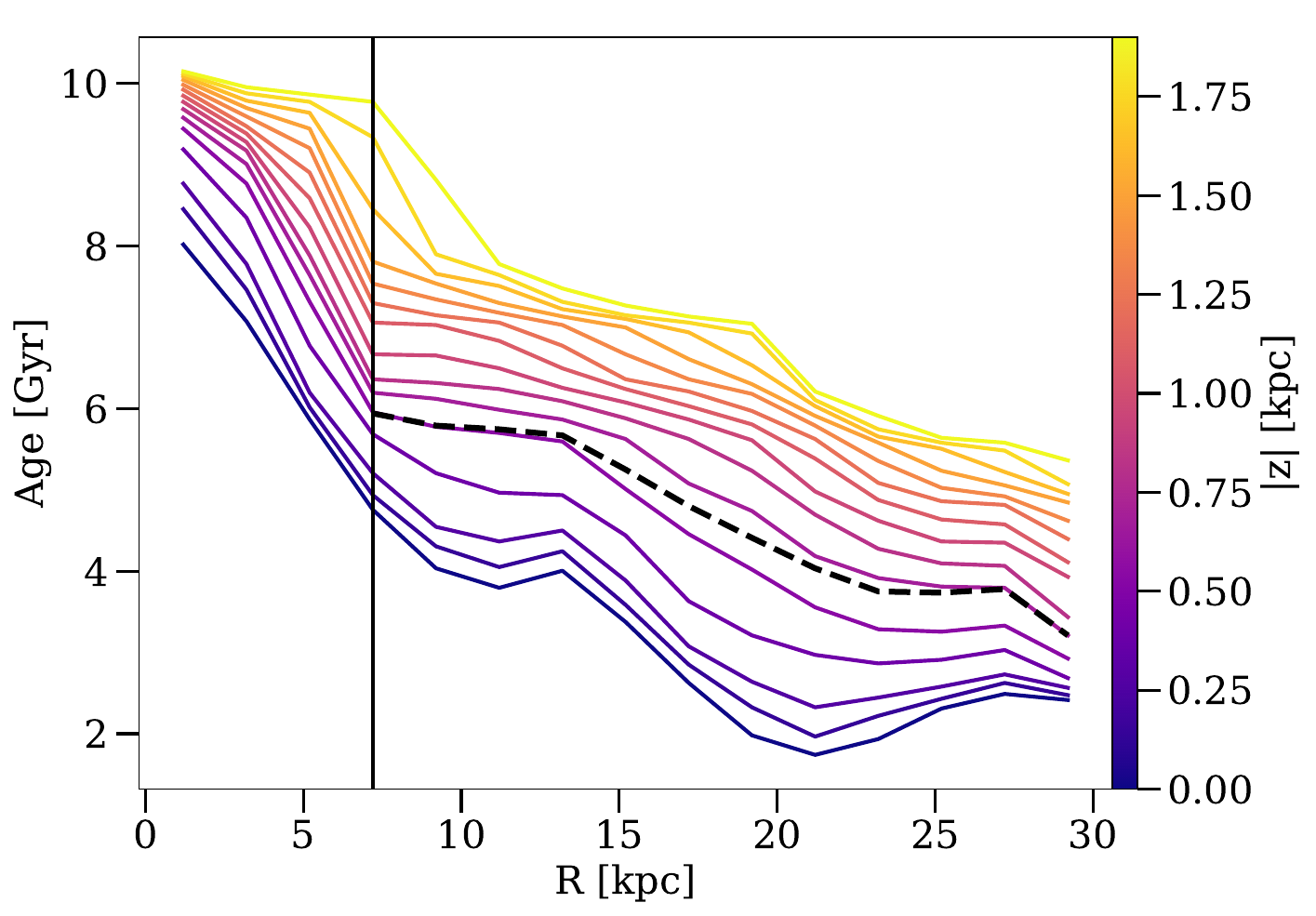}
\caption{Case 1a, illustrated by g92: both MAPs and thick disc flare minimally. \textit{Top left}: values of the scale-height against radius for the thin disc (triangles), thick disc (squares), and MAPs (solid lines) colour-coded by age. MAPs from 9 to 11 Gyr and from 11 to 13 Gyr are binned together, every other MAP spans 0.5 Gyr. The vertical black line on the left represents where we consider the galactic disc starts based on our criteria from Section \ref{sec:radial_extent}. A second vertical black line represents the end of the thick disc extent for cases when it is below 100\% of $R_{25}$. \textit{Top right}: surface density against radius for every MAP, binned and colour-coded in the same way as the top left panel. The vertical black lines represent the same as in the top left panel.
\textit{Bottom left}: surface density against scale-height for every MAP, colour-coded by age, at the middle of the disc. The black line is the histogram of the surface density, with 15 bins over the range of scale-heights. \textit{Bottom right}: median stellar age against radius colour-coded by height above the mid-plane. The dashed black line represents the age profile following the scale-heights of the thick disc. The vertical black lines represent the same as in the top left panel}
\centering
\label{fig:flatMAPflatDISC}
\end{figure*}

In this section, we will explain how this range of flaring for the thick discs is possible due to the connection between the flaring of the thick disc and the flaring of MAPs. As \cite{Minchev2015ONDISKS} explained, there are three main factors that come into play: 1) the level of flaring of the different MAPs, 2) their individual contribution to the total surface density, and 3) how these two change with radius. The latter is strongly connected with the inside-out formation of the disc. Non-flaring MAPs will not contribute to the flaring of the thick disc anywhere, rather, they will tend to minimise the flaring of the thick disc especially at those radii where their contribution to the total disc's surface density is more significant. Likewise, flaring MAPs that have a low surface density at the radii they flare, will not contribute either to the flaring of the thick disc. On the contrary, flaring MAPs with a substantial contribution to the total surface density at the radii they flare will drive thick disc flaring. Since the surface density contribution changes with radius for every MAP, the global flaring will be driven by different MAPs at different parts of the disc. 

\begin{figure*}
\centering
\includegraphics[scale=0.35]{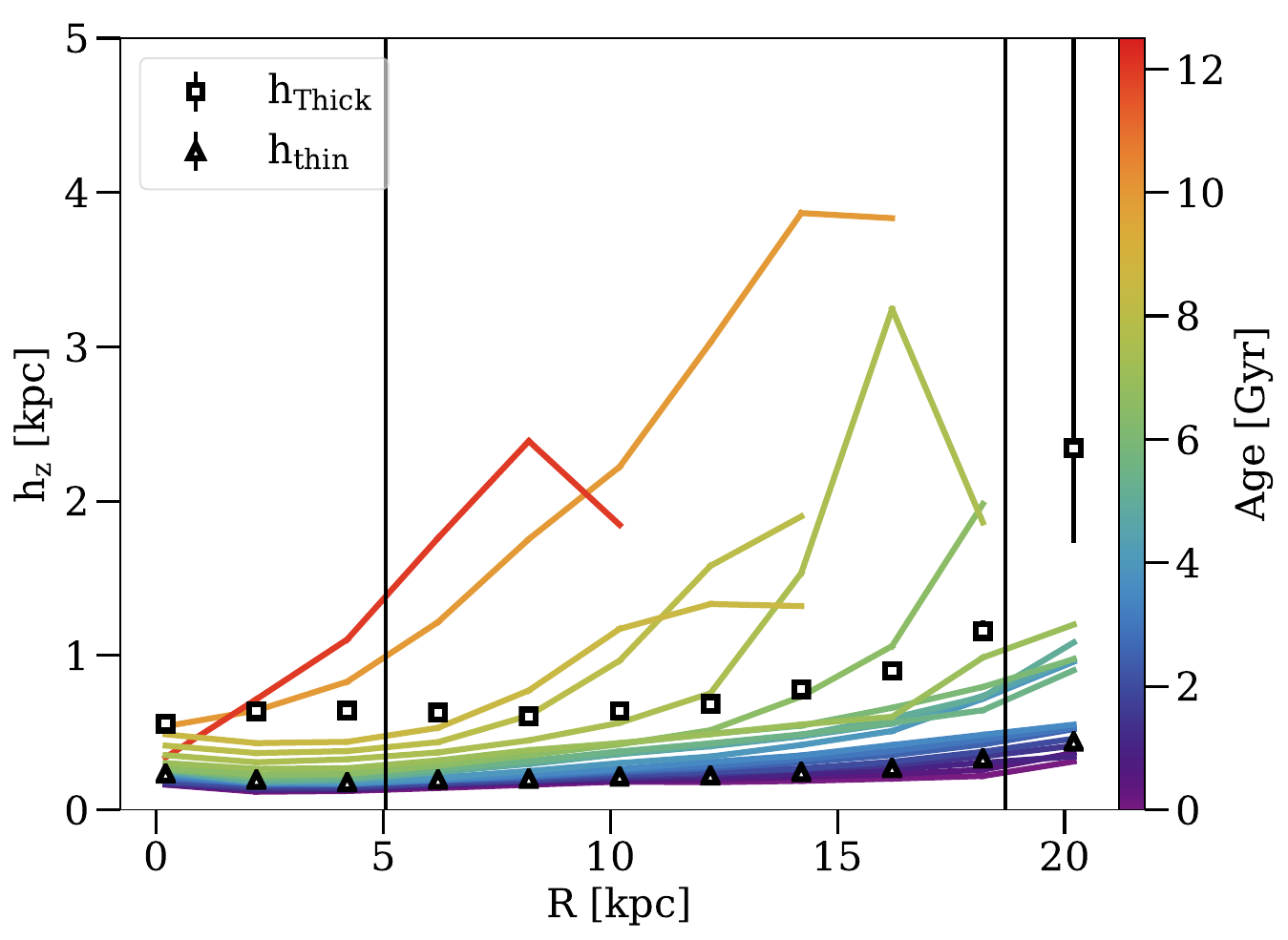}
\includegraphics[scale=0.35]{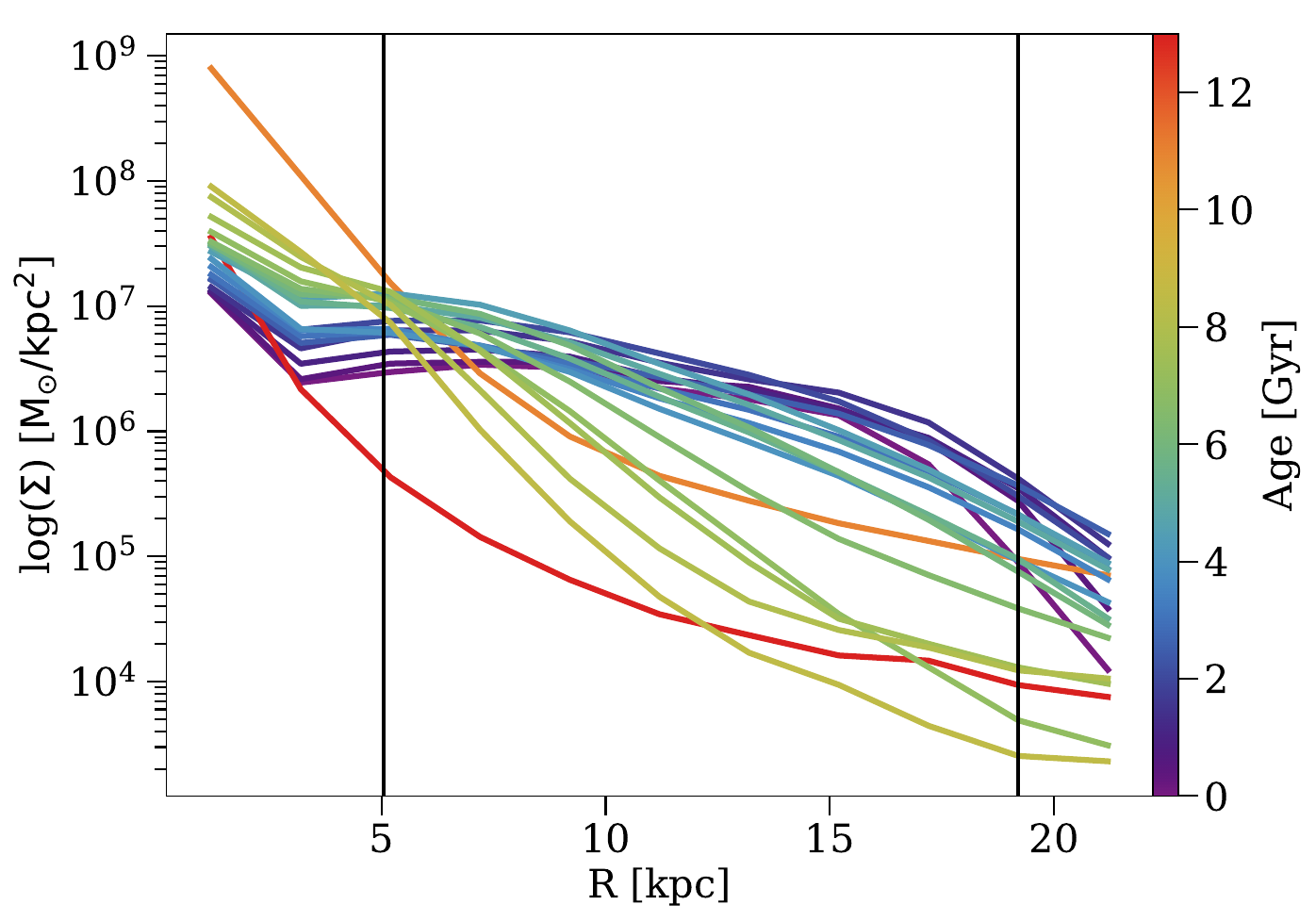}
\includegraphics[scale=0.35]{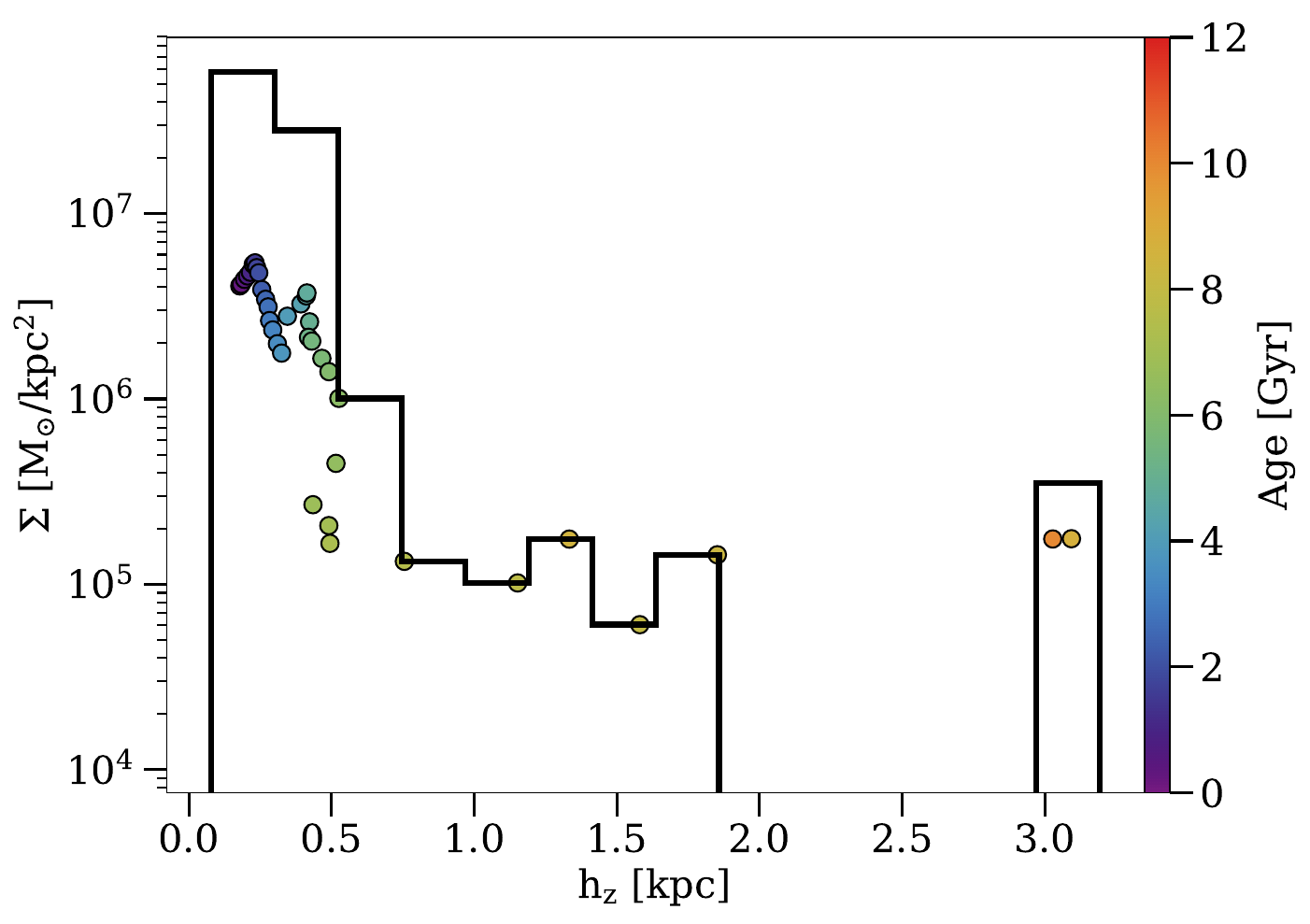}
\includegraphics[scale=0.35]{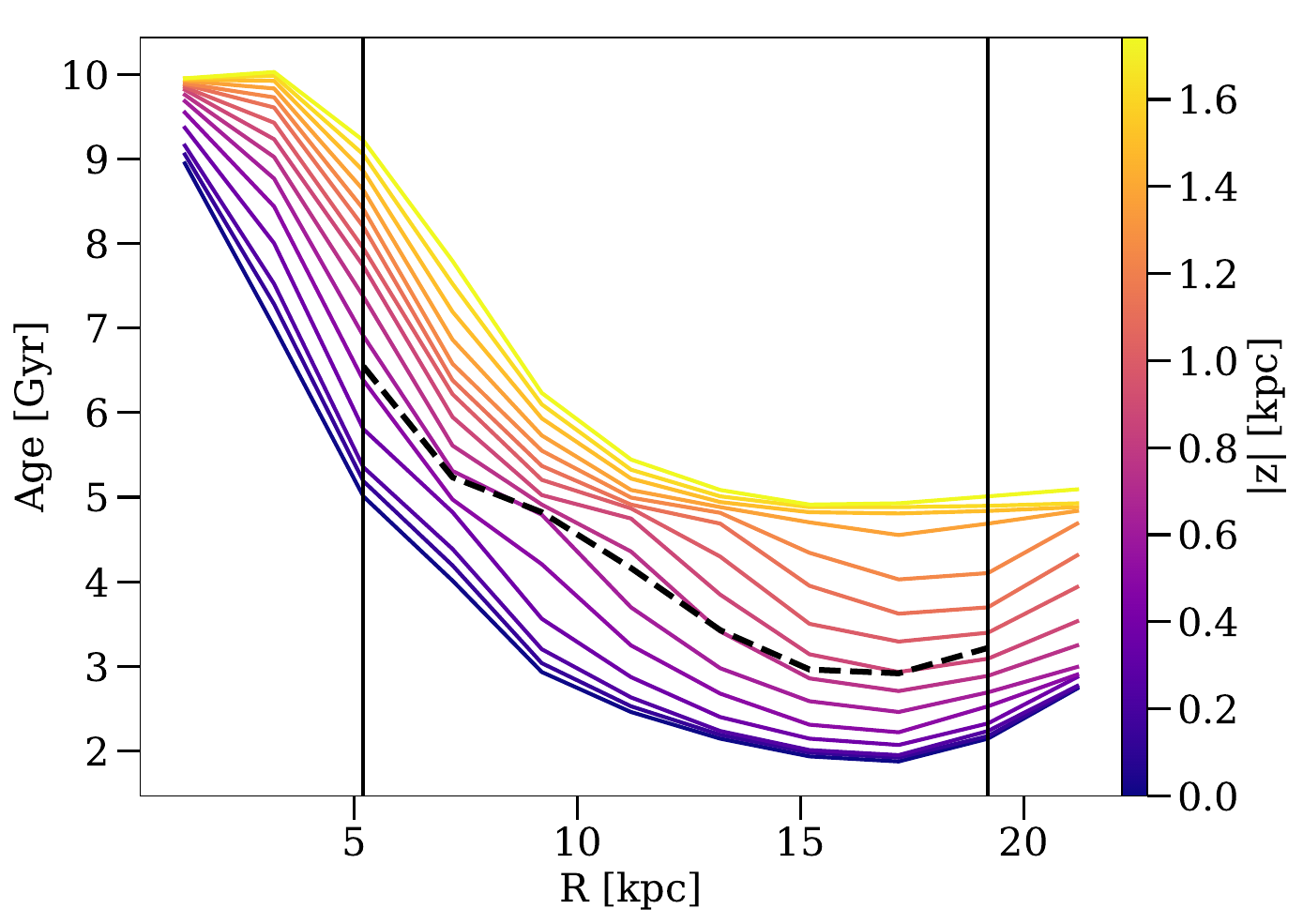}
\caption{Case 1b, illustrated by g47: flaring MAPs and minimally flared thick disc. Panels represent the same as in Fig. \ref{fig:flatMAPflatDISC}.}
\centering
\label{fig:flaredMAPflatDISC}
\end{figure*}

\subsection{General structure of MAPs} \label{sec:generalstructureMAPs}

To illustrate the full range of possible configurations for MAPs and global thin and thick discs, we pick five representative galaxies from our sample and show them in Figs. \ref{fig:flatMAPflatDISC} to \ref{fig:flaredMAPflaredDISCstack}. In the upper left panels of these figures, the scale-heights of the thin disc, thick disc, and every MAP, computed as explained in Section \ref{sec:scaleheight_fits}, are represented as a function of radius and colour-coded by age. 
Although we show only five examples, the reader can already see some general trends that hold in every galaxy within our sample. Older MAPs --- especially older than 9 Gyr --- tend to be concentrated in the inner part of the galaxy, i.e. within $R_{\mathrm{inner}}$ (the radius where the disc starts, see Section \ref{sec:radial_extent}). They also flare strongly with radius, with already large scale-heights at the beginning of the disc and very low surface density values. This would imply that these MAPs' contribution to the global disc flaring should be negligible. 
In order to be sure, we re-computed the thick disc slopes $\nabla_{\mathrm{Thick}}$ excluding all stellar particles 9 Gyr old and older. We find extremely small changes for  $\nabla _{\mathrm{Thick}}$ in the vast majority of the galaxies in our sample, which confirms the negligible influence of the oldest stars on the flaring of the thick disc.
This does not mean that MAPs older than 9 Gyr do not have an influence over the global shape of the thick disc.
When excluding all stellar particles older than 9 Gyr, we find that the scale-heights of the global thick disc have lower values by 25~\% on average.
Within certain variability, this effect has the same intensity all over the disc. 
This explains why the flaring level is unaffected by these MAPs.  
In other words, MAPs older than 9 Gyr thicken the disc and increase the scale-heights of the global thick disc uniformly but do not have an influence on the scale-heights' gradient. 

The youngest MAPs --- younger than 3 Gyr --- generally dominate the surface density of the disc, especially towards the outskirts. 
They are normally quite flat, show very little flaring compared to older MAPs, and their scale-heights are too low to affect the shape of the thick disc. 
Therefore, generally speaking, MAPs younger than 3 Gyr old dictate the flaring of the thin disc. This can be seen in the upper left panels of Figs. \ref{fig:flatMAPflatDISC} to \ref{fig:flaredMAPflaredDISCstack}, where the thin discs' scale-heights follow very tightly the flaring of these MAPs. Based on all this, we can say that MAPs older than 9 Gyr and younger than 3 Gyr old contribute little or nothing to thick disc flaring.
Intermediate age MAPs --- from 3 to 9 Gyr roughly --- are the ones which exhibit the most varied behaviours, from flaring minimally to dramatically. 
In some galaxies, the scale-heights and flaring level of MAPs increase gradually with age.
In others, there are gaps between particular MAPs, or quite a few MAPs have roughly the same scale-heights. \cite{Martig2014a} found that these gaps are linked to the galaxy merger history. 
The different flaring scenarios showed by the intermediate age MAPs, together with their contribution to the total surface density of the disc at every radii, will dictate the flaring level of the thick disc.

\subsection{Flaring of MAPs and flaring of the thick disc} \label{sec:flaringmapsflaringthickdisc}

Generally speaking, we establish two main scenarios based on whether the thick disc is flat or flared.
We would like to stress that flat and flared thick discs are not two completely distinct categories, and the galaxies in our sample exhibit a continuous spectrum between the two. Based on visual criterion, we pick a value of $\mathrm{\nabla_{Thick}}$ around 0.03 to split galaxies into flat or flared, keeping in mind that this number is somewhat arbitrary.
We also would like to remind the reader that flat does not mean an absolute absence of flaring but rather very small or minimal flaring. 
In this subsection, we first pick five galaxies which are representative of the different ways MAPs create thick discs and explain how the different mechanisms come into play. Afterwards, we analyse the whole galaxy sample.

\subsubsection{The five cases}
\begin{figure*}
\centering
\includegraphics[scale=0.35]{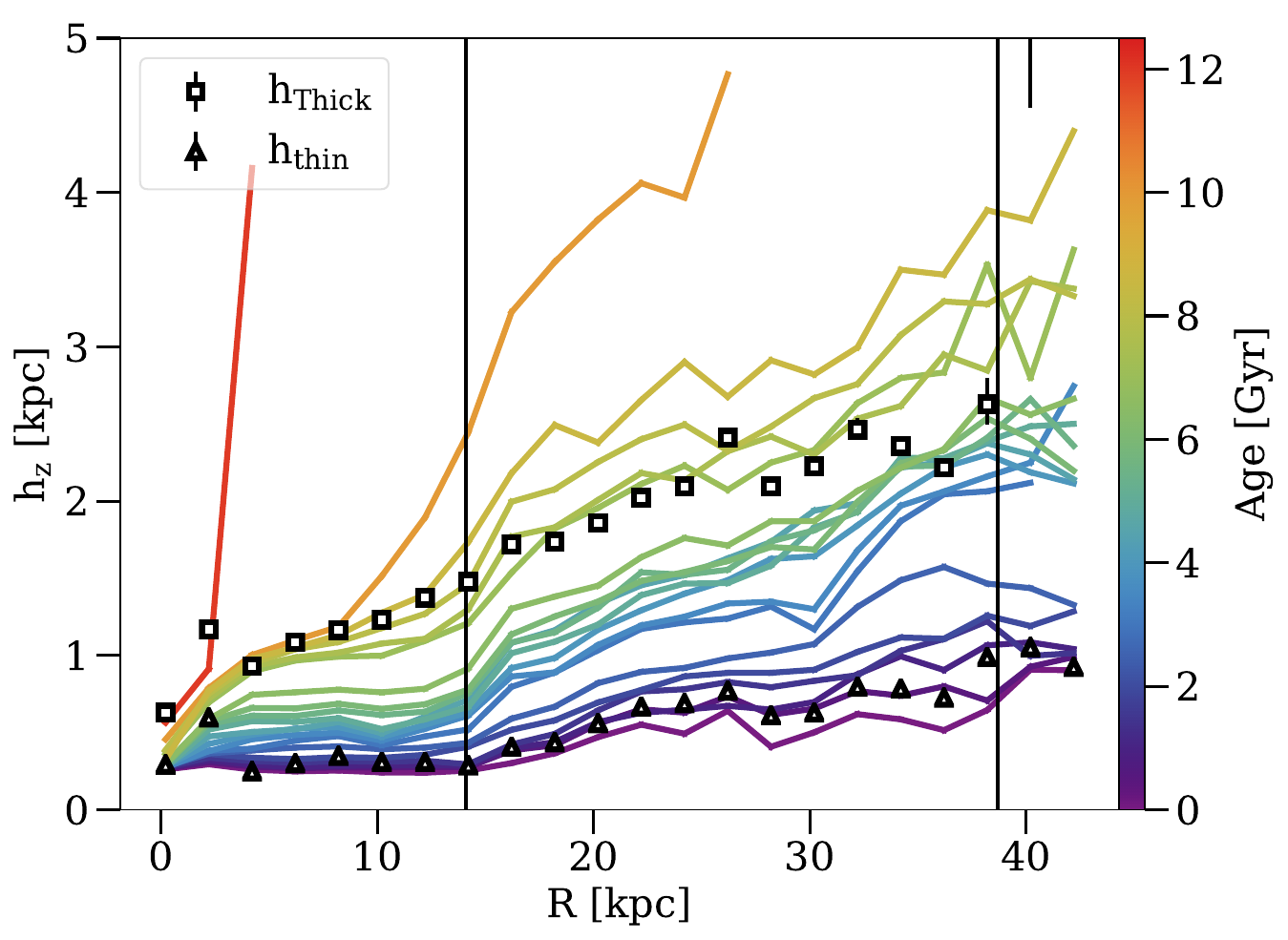}
\includegraphics[scale=0.35]{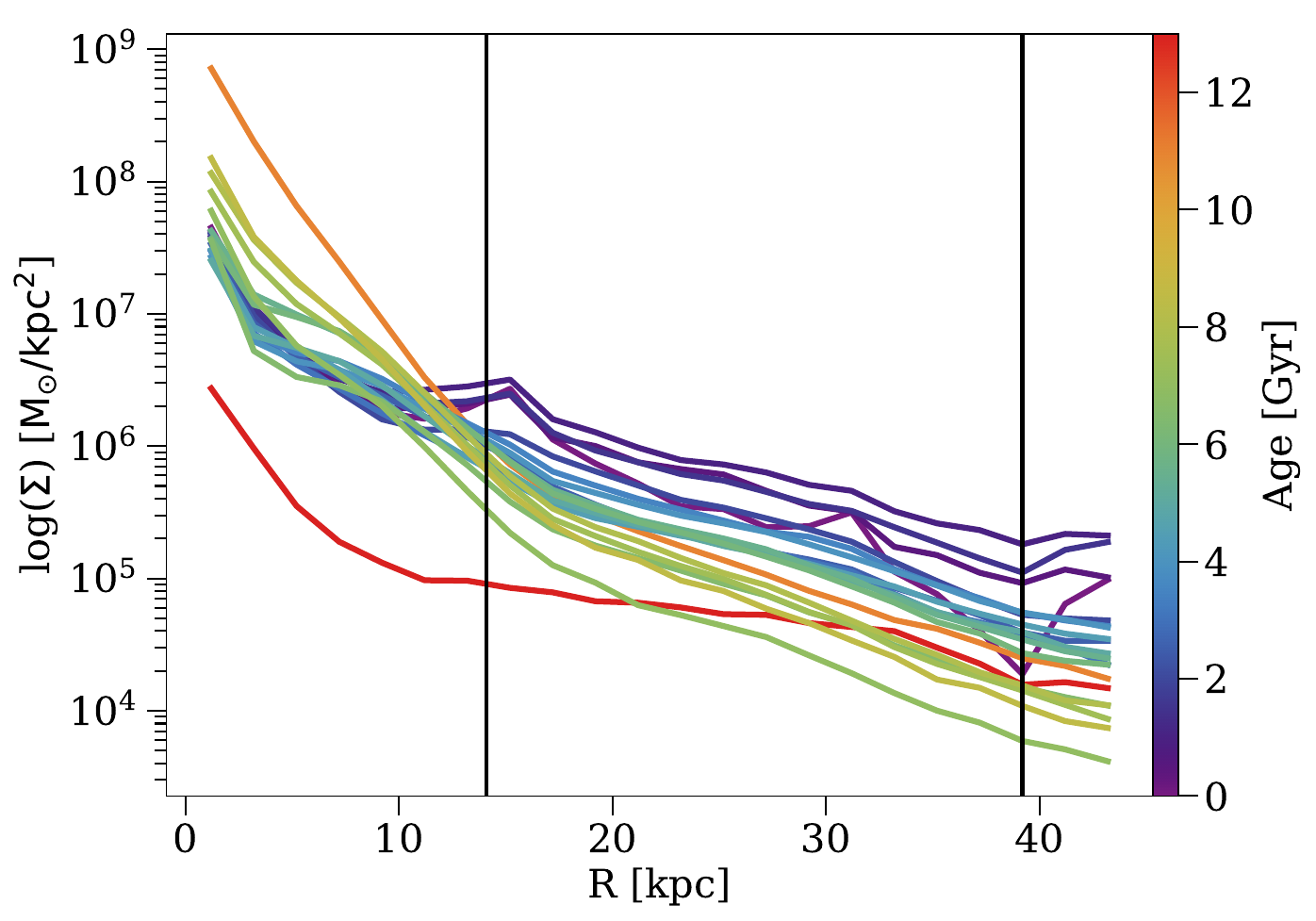}
\includegraphics[scale=0.35]{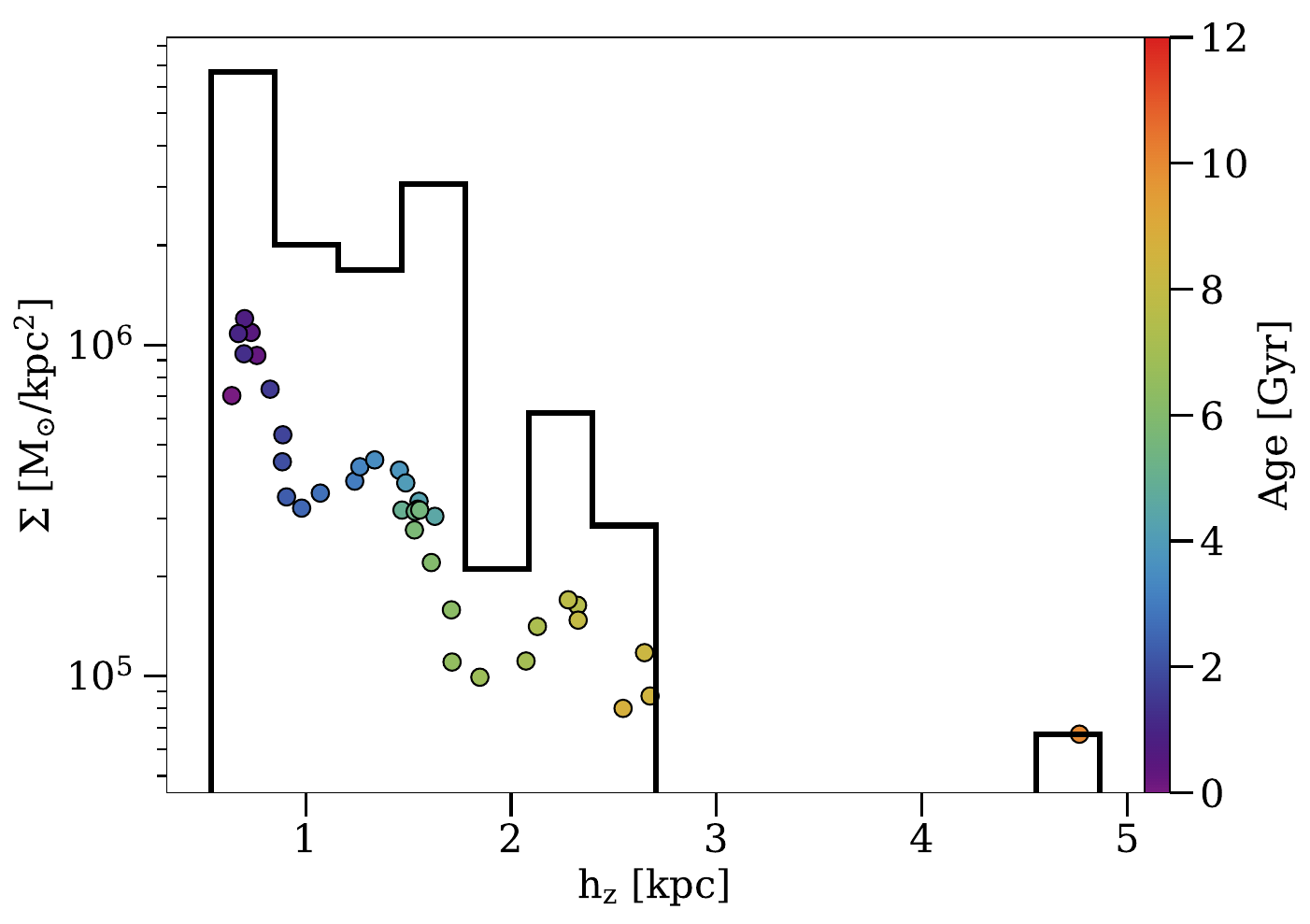}
\includegraphics[scale=0.35]{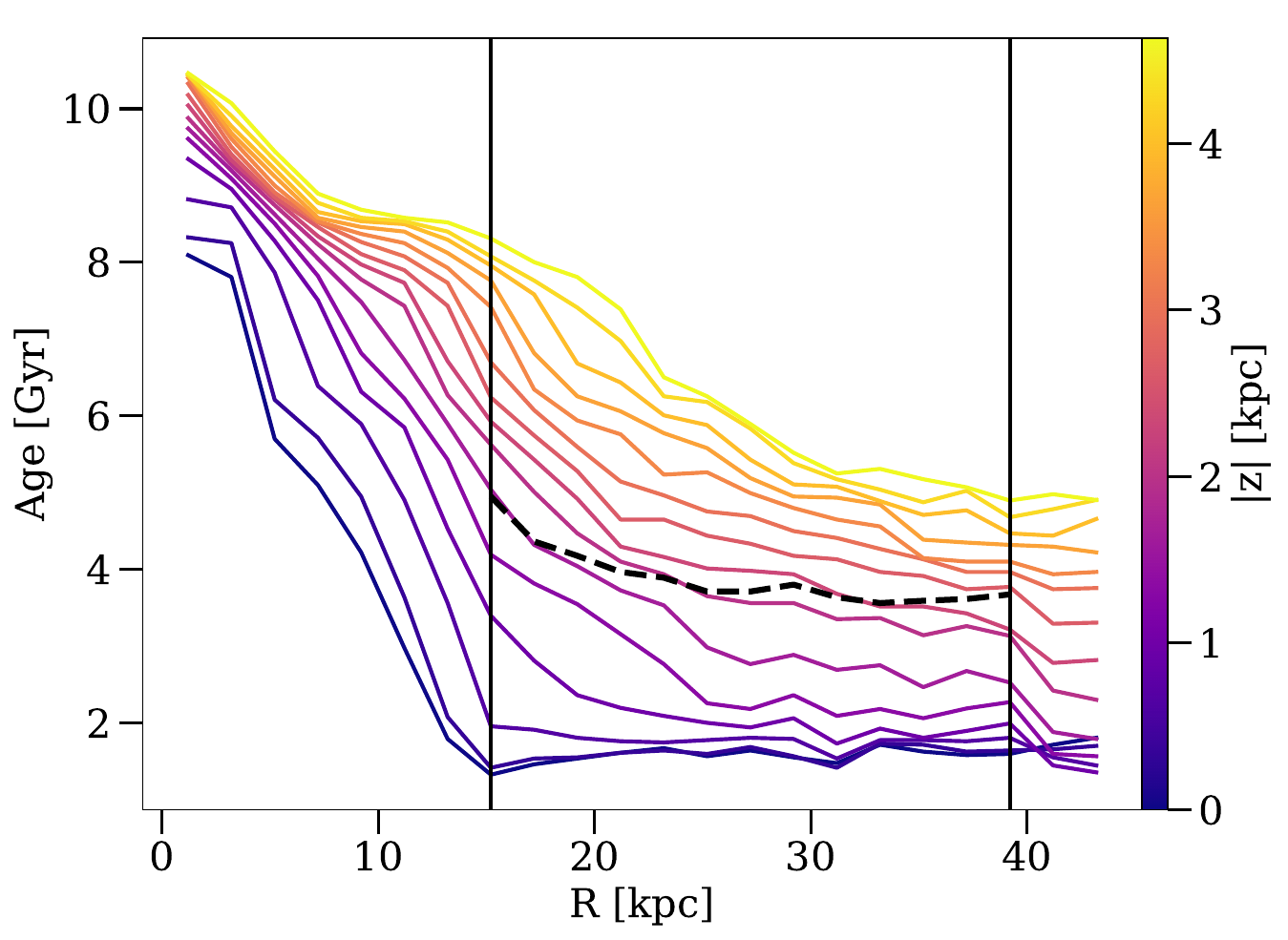}
\caption{Case 2a, illustrated by g36: flaring MAPs and flaring thick disc. Panels represent the same as in Fig. \ref{fig:flatMAPflatDISC}}.
\centering
\label{fig:flaredMAPflaredDISC}
\end{figure*}

\begin{figure*}
\centering
\includegraphics[scale=0.35]{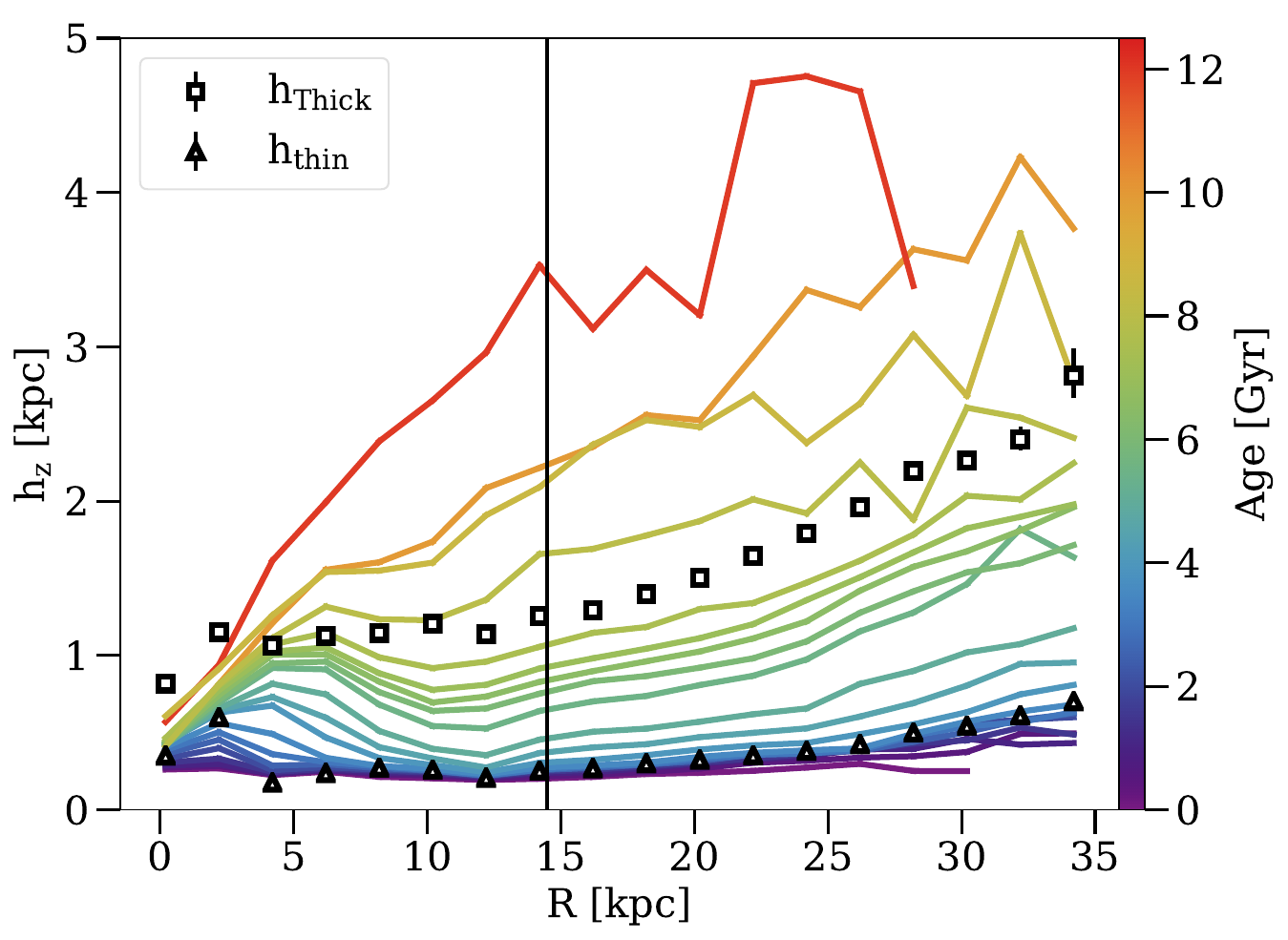}
\includegraphics[scale=0.35]{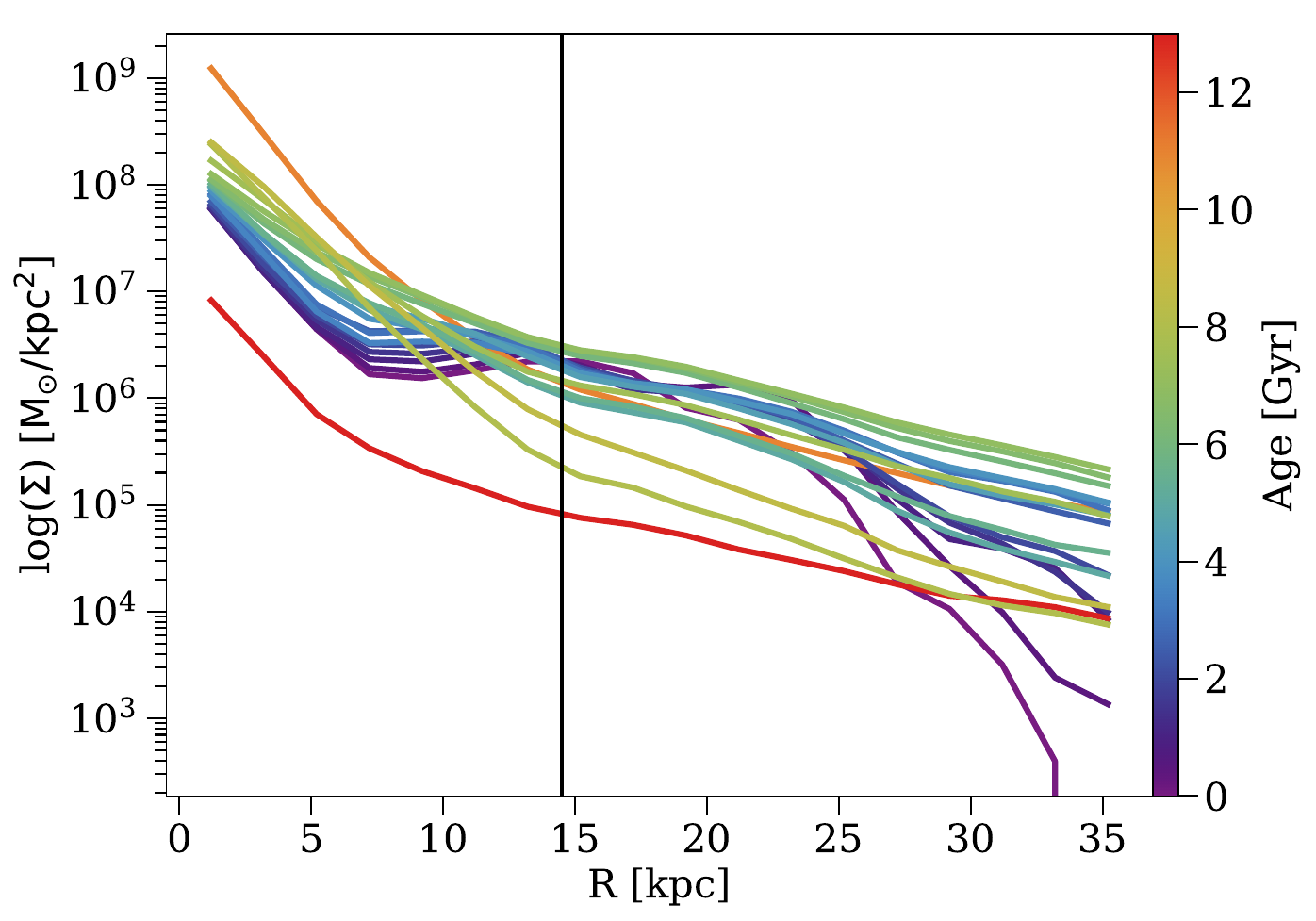}
\includegraphics[scale=0.35]{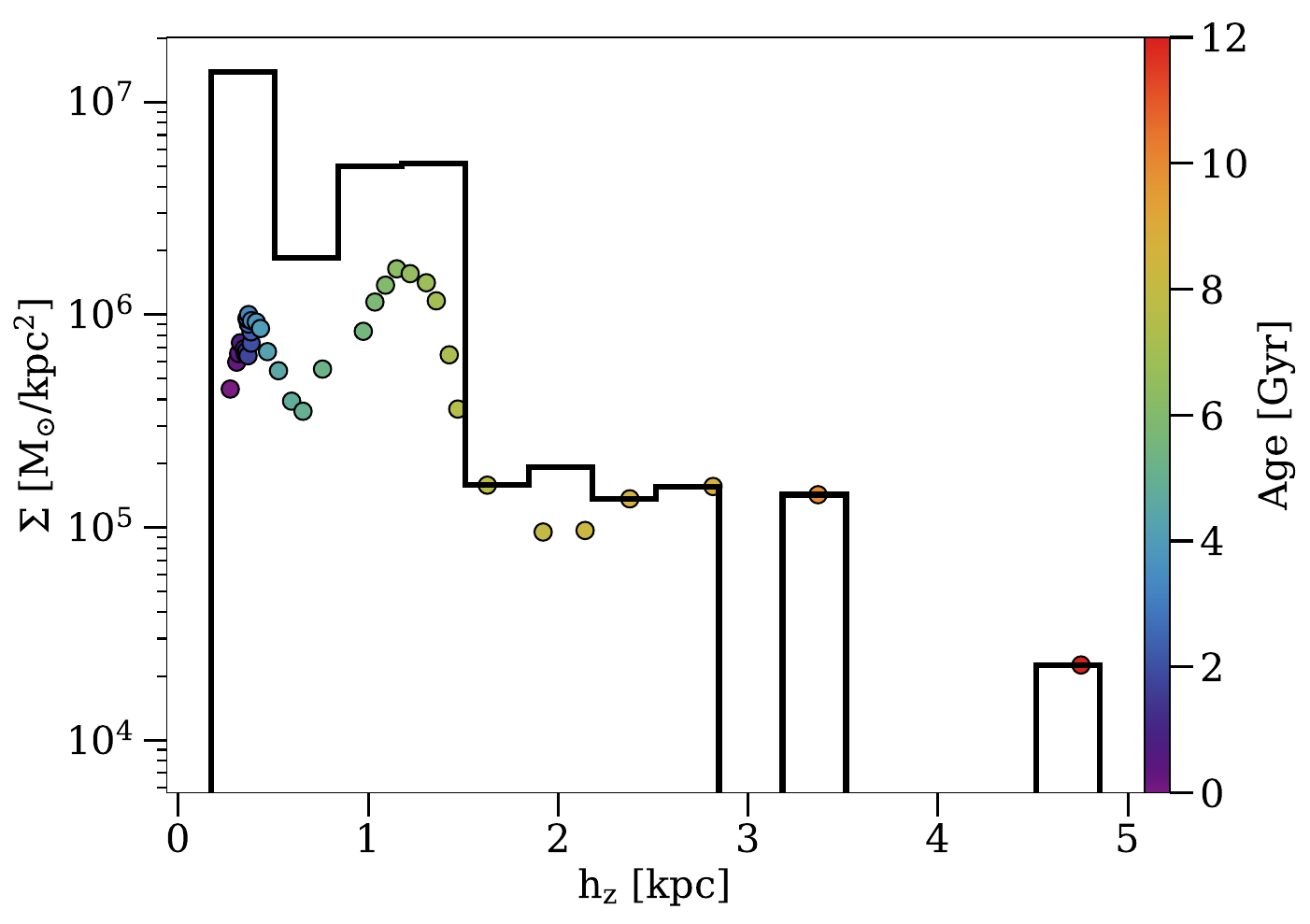}
\includegraphics[scale=0.35]{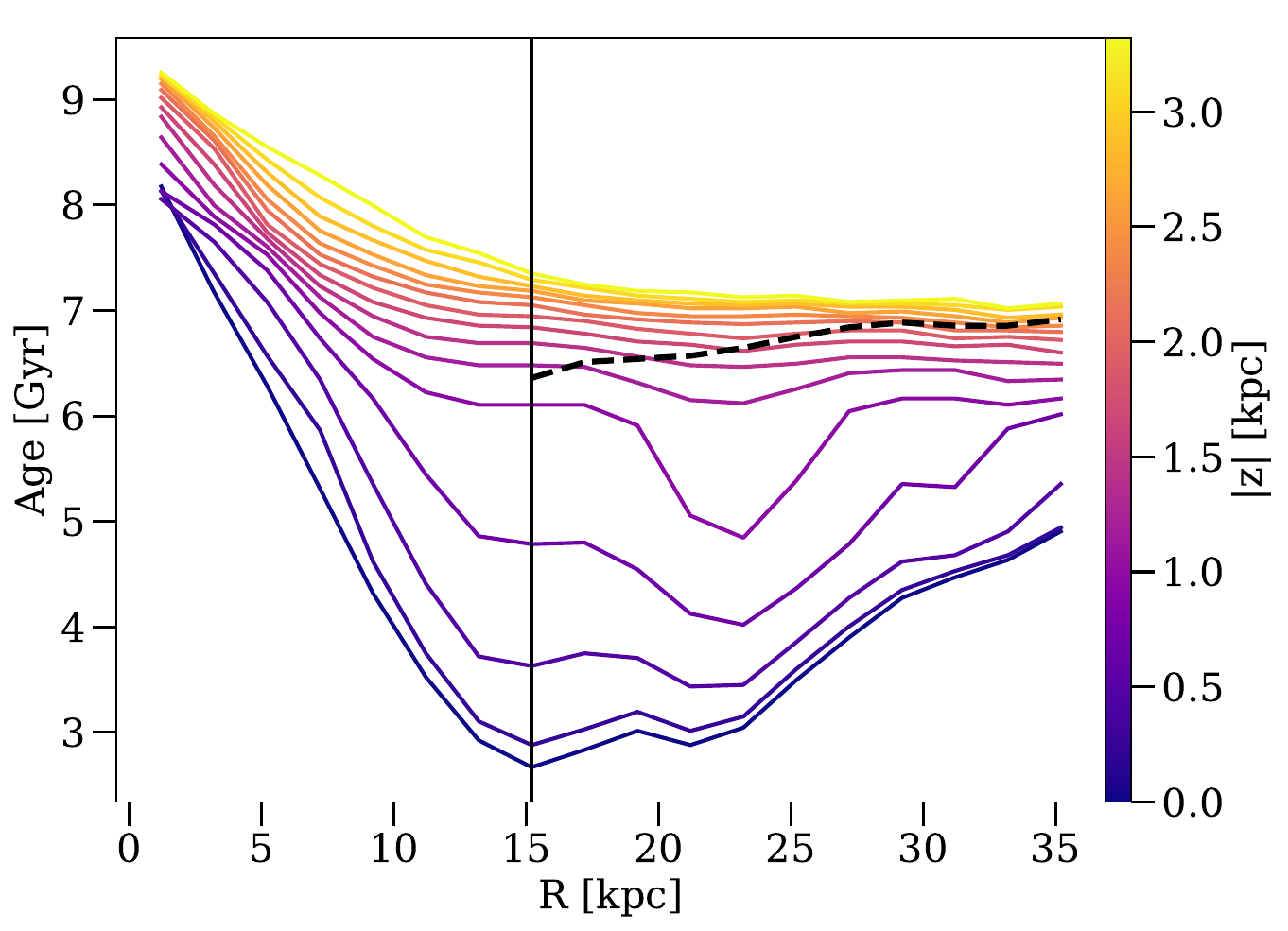}
\caption{Case 2b, illustrated by g39: a case 2a with the particularity of having a few flaring MAPs dominating the surface density at all radii. The thick disc's scale-height follows those MAPs' flaring. Panels represent the same as in Fig. \ref{fig:flatMAPflatDISC}.}
\centering
\label{fig:flaredMAPflaredDISCextreme}
\end{figure*}

\begin{figure*}
\centering
\includegraphics[scale=0.35]{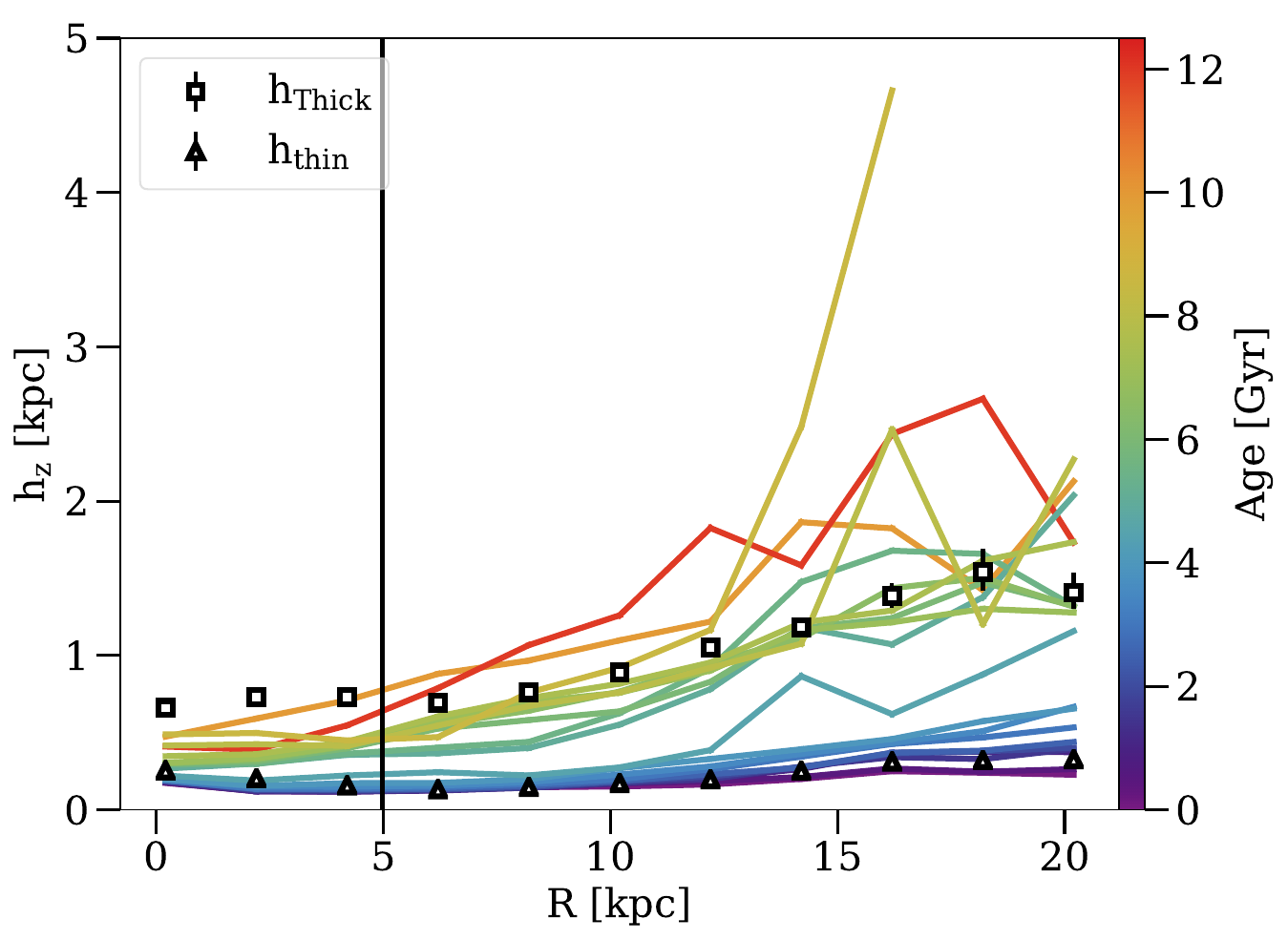}
\includegraphics[scale=0.35]{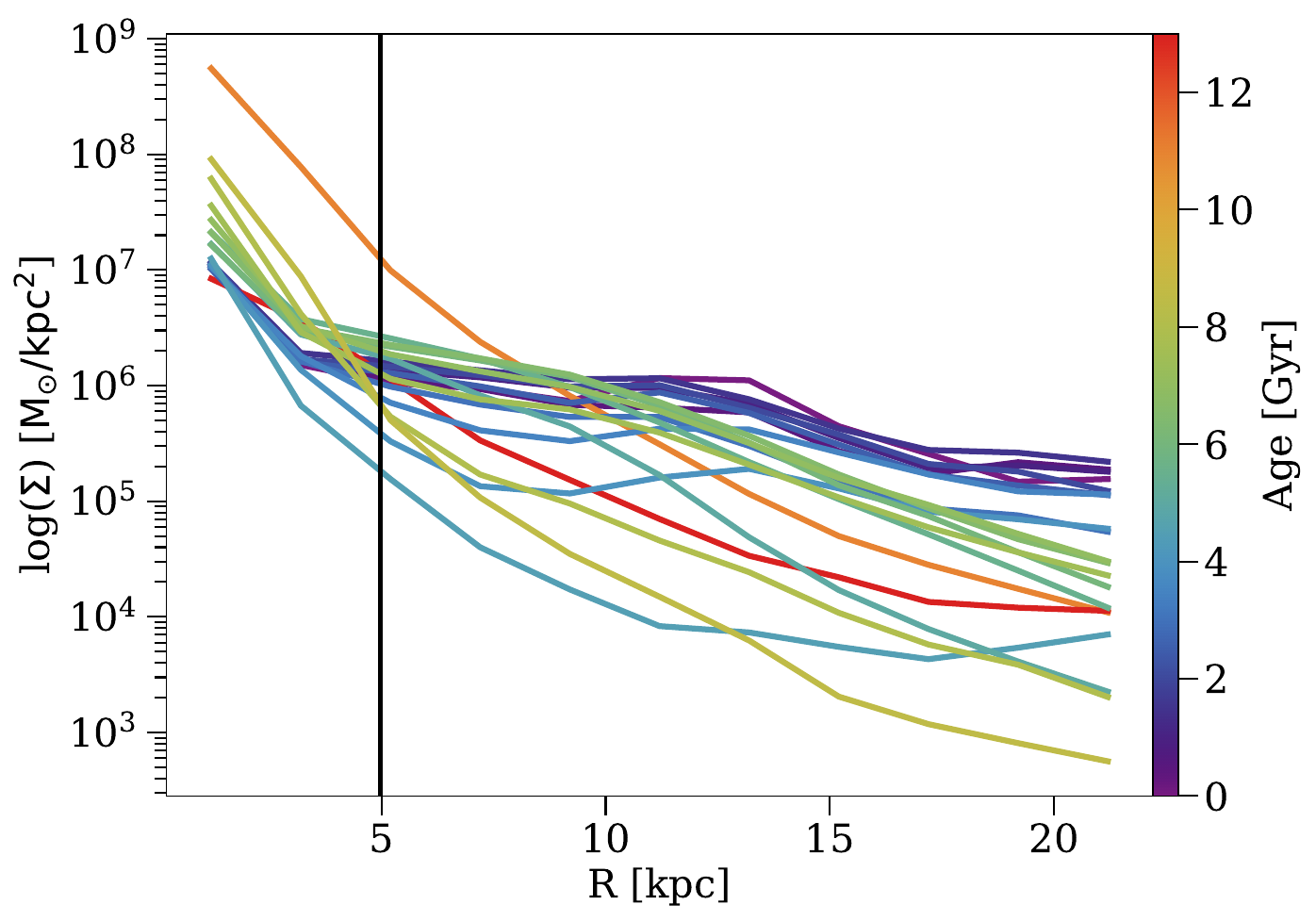}
\includegraphics[scale=0.35]{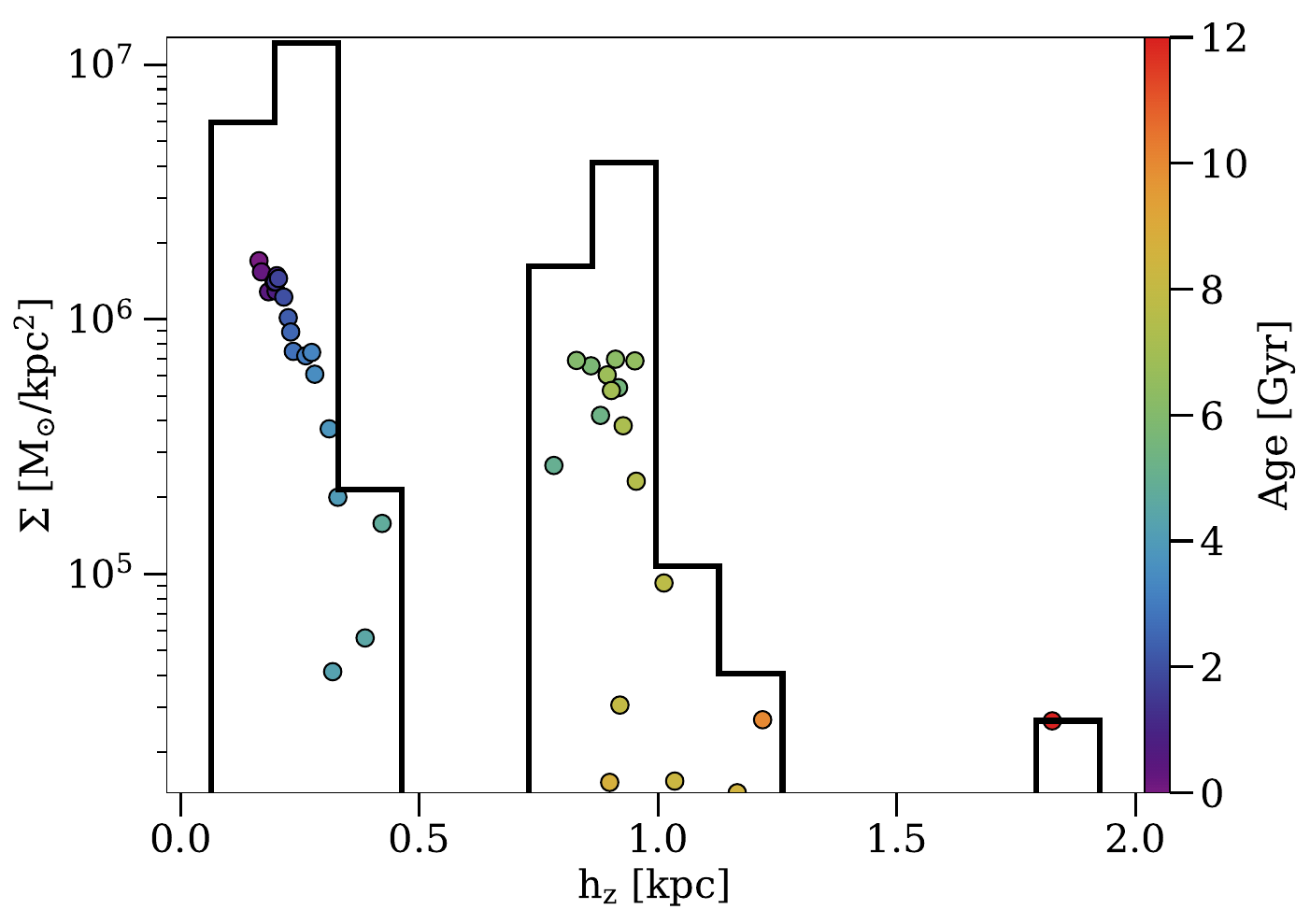}
\includegraphics[scale=0.35]{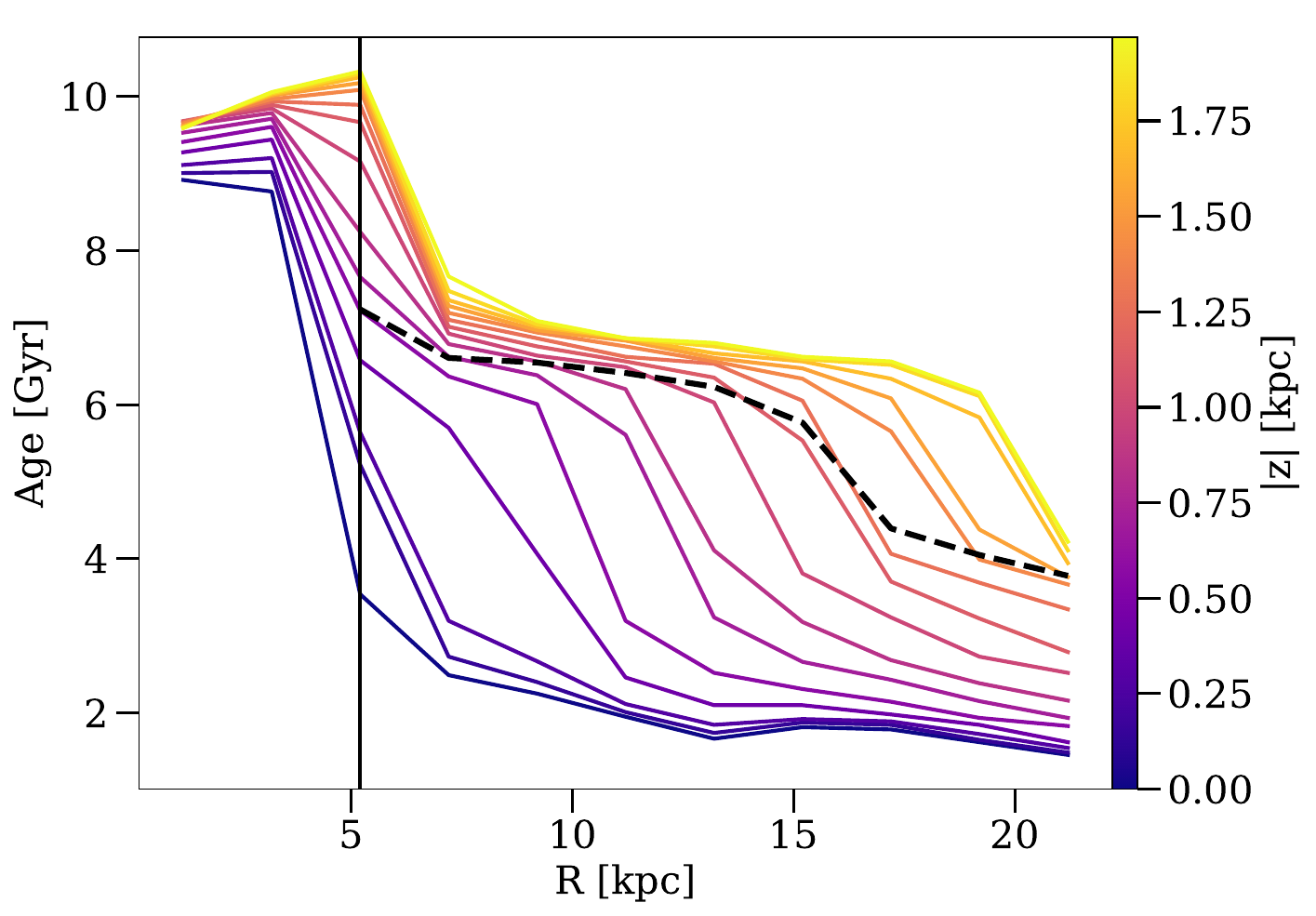}
\caption{{Case 2c, illustrated by g102: a case 2a with the particularity of a group of flaring MAPs having similar scale-heights and the thick disc's scale-height following those MAPs' flaring. Panels represent the same as in Fig. \ref{fig:flatMAPflatDISC}}}
\centering
\label{fig:flaredMAPflaredDISCstack}
\end{figure*}

Out of the five representative galaxies we have chosen, cases \textbf{1a} and \textbf{1b} in Figs. \ref{fig:flatMAPflatDISC} and \ref{fig:flaredMAPflatDISC} represent two possible ways to obtain a flat thick disc, whereas cases \textbf{2a}, \textbf{2b}, and \textbf{2c} in Figs. \ref{fig:flaredMAPflaredDISC}, \ref{fig:flaredMAPflaredDISCextreme}, and \ref{fig:flaredMAPflaredDISCstack} represent three possible ways to obtain a flared thick disc. 

Galaxy g92 in Fig. \ref{fig:flatMAPflatDISC} represents our \textbf{case 1a}, which is when all or most MAPs are flat or flare minimally. 
Galaxies in this category have MAPs with scale-heights that do not increase much in absolute value as a function of radius. This can still sometimes correspond to relatively large fractional increases.
For instance, in g92, MAPs between 5 and 8 Gyr almost double their scale-height values from $R_{\mathrm{inner}}$ to $R_{\mathrm{outer}}$, but in absolute values this represent an increase of less than 0.5 kpc over a 15 kpc disc radius.

The values of the global thick disc's scale-heights do not follow one particular MAP's scale-heights, but a combination of all MAPs' scale-heights following the mechanism proposed by \citealp{Minchev2015ONDISKS}. Still, there is no stellar population that could drive strong thick disc flaring as all of them are practically flat. 
In the upper right panel of Fig. \ref{fig:flatMAPflatDISC}, we show the stellar surface density against radius for every MAP (only including particles up to a height of three times $h_{\mathrm{scale}}$).
We find that different MAPs contribute at different radii, in particular the inner disc is dominated by older populations and the outer disc by younger populations, reflecting the inside-out formation process of the disc.
This combination of MAPs with different radial surface density profiles results in the scale-height values of the flat thick disc.
Also, this succession of dominant MAPs in the radial direction creates an age gradient in the thick disc as will be described below in Section \ref{sec:agegradients}.
In other words, there are no MAPs with significant flaring and significant surface density to drive thick disc flaring. Yet, the interplay between flaring, surface density, and radius determines the global thick disc's scale-height values and creates an age gradient. In our sample we find 6 galaxies in this category: g31, g82, g86, g92, g128, and g148.

The other way to make a flat thick disc is our \textbf{case 1b}, when some MAPs flare considerably more than the rest, but they have low surface density compared to the rest of MAPs at the flaring radii, like for example in g47 illustrated in Fig. \ref{fig:flaredMAPflatDISC}. 
In the upper left panel, it can be seen that MAPs above 6 Gyr old flare considerably towards the outskirts which could potentially make the global thick disc flare. Nevertheless, when we look at how much those MAPs contribute to the surface density in the upper right panel of Fig. \ref{fig:flaredMAPflatDISC}, only MAPs younger than 5 Gyr old dominate the surface density in the outer part of the disc ---again, due to the inside-out formation of the disc. 
Therefore, the flaring MAPs are important for the scale-height values of the inner disc, but do not have much influence on the scale-height values of the outskirts, producing very subtle or no thick disc flaring. 
We find 6 galaxies in our sample that fit in this category: g37, g38, g47, g62, g83 and g106. 

Flared thick discs, on the other hand, are created when MAPs have a significant surface density at the radius they start flaring. In this scenario, the scale-height of the global thick disc will progressively follow the flare of different MAPs, depending on how their respective surface densities and flaring levels combine, which is a result of the inside-out formation process of the disc. Galaxy g36 in Fig. \ref{fig:flaredMAPflaredDISC} represents our \textbf{case 2a}.
As can be seen in the upper left panel, MAPs older than 3 Gyr old flare considerably. There are even some gaps between the MAPs' scale-heights which are connected to mergers as discussed in \cite{Martig2014a}. Similarly to case 1a and 1b represented in Figs. \ref{fig:flatMAPflatDISC} and \ref{fig:flaredMAPflatDISC}, as the surface density of the older MAPs gets lower with radius both in absolute values and relatively to younger MAPs, shown in the upper right panel of Fig. \ref{fig:flaredMAPflaredDISC}, the thick disc's scale-heights move across the gap between the MAPs' scale-heights and follow those of the younger components by the outskirts of the disc. 
This represents the most common flared thick disc scenario in our sample with 9 galaxies fitting into this category: g22, g36, g56, g57, g72, g124, g126, g136, and g147. 

Cases \textbf{2b} and \textbf{2c} represent two particular scenarios of the mechanisms operating in case \textbf{2a}. In \textbf{case 2b}, only one or few flaring MAPs dominate the surface density all the way throughout the disc. 
If that is the case, then the thick disc's scale-height will follow closely the flaring of that particular MAP/MAPs. An example is g39, represented in Fig. \ref{fig:flaredMAPflaredDISCextreme}. In the upper left panel of Fig. \ref{fig:flaredMAPflaredDISCextreme}, the thick disc's scale-height follows MAPs around 7-7.5 Gyr old. If we look at the upper right panel of Fig. \ref{fig:flaredMAPflaredDISCextreme}, we see that out of the MAPs contributing to the thick disc, MAPs between 6 and 7.5 Gyr old dominate by far the surface density at all radii. Therefore, these MAPs are going to dictate the flaring of the thick disc and its scale-heights are not going to go across multiple MAPs. Only 2 galaxies in our sample fit into this category: g39 and g44. 

Finally, galaxy g102 represented in Fig. \ref{fig:flaredMAPflaredDISCstack} shows our last flared thick disc scenario, \textbf{case 2c}. Although both MAPs and global thick disc are flared, instead of MAPs increasing their flaring and scale-heights progressively with age like cases 2a and 2b, a big group of MAPs have the same scale-heights. 
This can be seen in the upper left panel of Fig. \ref{fig:flaredMAPflaredDISCstack}. The upper right panel of Fig. \ref{fig:flaredMAPflaredDISCstack} shows that in the thick disc, the surface density is dominated by different MAPs as a function of radius because of the inside-out formation  process of the disc. Nevertheless, since all these MAPs have roughly the same values of the scale-height at all radii, the thick disc's scale-heights will follow this group of MAPs. 4 galaxies from our sample fit into this category: g45, g48, g59, and g102.

\begin{figure}
    \centering
    \includegraphics[width=\columnwidth]{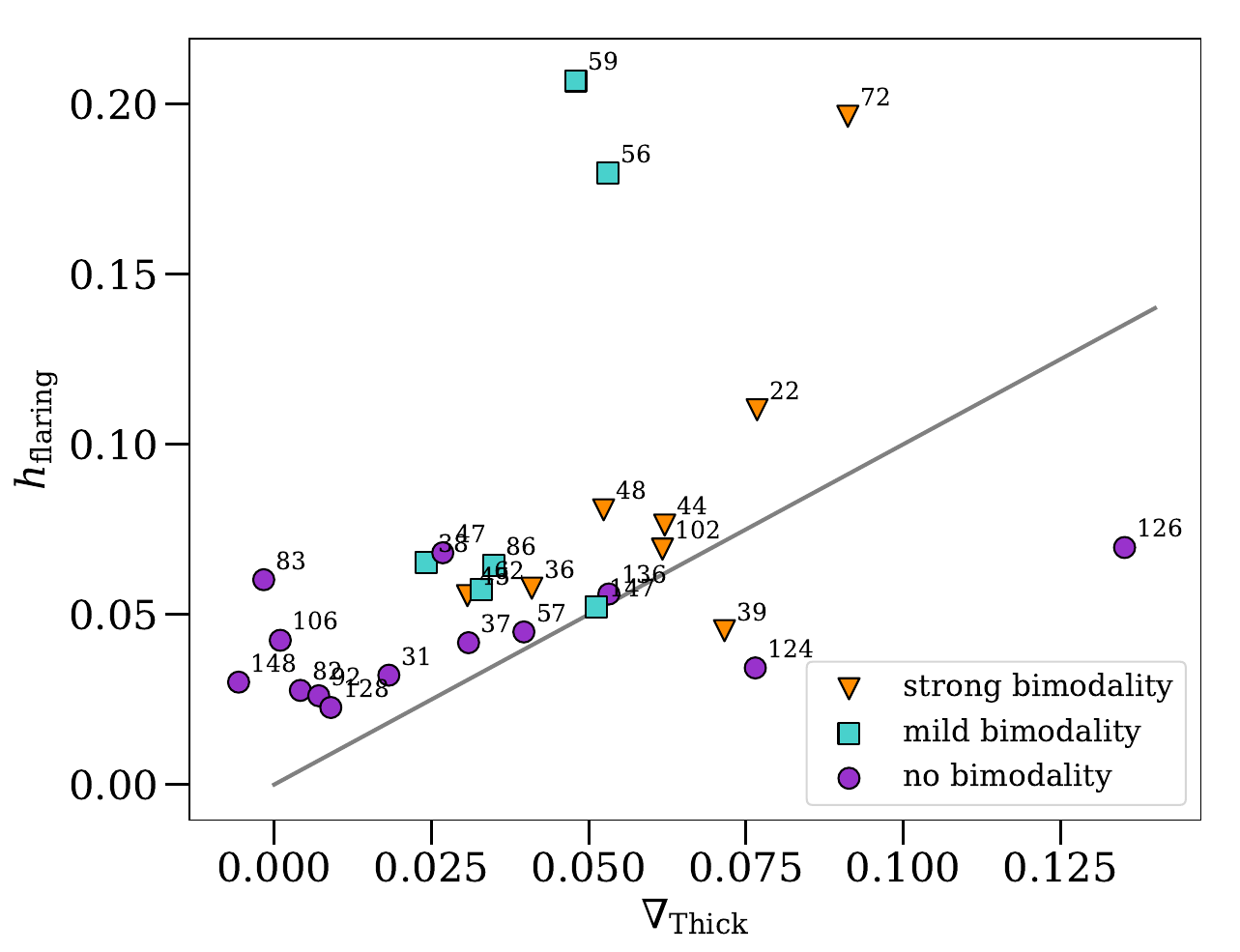}
    \caption{Mass averaged slope of all MAPs against thick disc slope. The marker shapes represent the bimodality of thin/thick disc: triangles represent a bimodal thin/thick structure, circles represent a continuum structure, and squares represent an intermediate case between continuum and bimodality. A black line shows a 1:1 line for comparison purposes. For almost all galaxies, MAPs are always more flared than the thick disc. All flat thick disc galaxies show no bimodality between thin and thick discs and their MAPs flare less than their flared thick disc counterparts, most of which show a bimodal structure.}
    \label{fig:flaring factor}
\end{figure}

\subsection{Bimodal vs. continuous structure} \label{sec:bimodalVScontinuous}

Because of the different configurations of flaring MAPs and surface density changing in the radial direction,  in some galaxies the thin and the thick discs will be formed by different groups of stellar populations, hence being two distinct components, whereas in some others the stellar populations forming the thin and the thick discs will be much more mixed ---hence forming a thin-thick disc continuum. In the bottom left panel of Figs. \ref{fig:flatMAPflatDISC} to \ref{fig:flaredMAPflaredDISCstack}, the surface density of each MAP colour-coded by age is represented against the value of their scale-height at $R_{1/2}$.

On the one hand, a thin/thick disc bimodality scenario is one where the disc has two components with comparable surface densities and each component has only certain MAPs. Galaxies g39 and g102 in Fig. \ref{fig:flaredMAPflaredDISCextreme} and \ref{fig:flaredMAPflaredDISCstack} respectively are representative of this bimodality. In g102, the younger MAPs do not flare much and form the thin disc, whereas the older MAPs flare ---and have the same scale-heights--- and form the thick disc. Therefore, the thin and thick discs are clearly spatially divided and the stellar populations populating them are completely different regarding their age. Also, both disc components have a considerable amount of surface density. Galaxy g39 also shows a bimodality in its thin/thick disc. Contrary to g102, there are no big gaps between the MAPs' scale-heights and  there is no group of MAPs with the same scale-height. However, a group of a few MAPs in the thick disc is so much more dominant in surface density than the rest, that this group creates a distinct massive thick disc component. We find a total of 8 galaxies in our sample that show clear thin/thick disc bimodality.

On the other hand, a continuum scenario is one where the transition between MAPs forming the thin disc and those of the thick disc is smooth, without clear massive distinct components. Galaxies g92 and g47 in Figs. \ref{fig:flatMAPflatDISC} and \ref{fig:flaredMAPflatDISC} respectively are good examples: the scale-heights of MAPs increase progressively and no MAP dominates significantly over the others in terms of surface density. 
We find a total of 13 galaxies in our sample show a continuous thin/thick disc structure.

Finally we find 6 galaxies in our sample that we classify as intermediate cases. These galaxies do not have a single gap in their MAPs' scale-heights but a few of them. However, because of the changing surface density with radius, sometimes the global thick disc's scale-heights are above a particular gap and sometimes below depending on radius.
As a consequence, there is no clear group of MAPs which dictate the thick disc's scale-heights throughout the disc. Galaxy g36 in Fig. \ref{fig:flaredMAPflaredDISC} is a good example of the intermediate category.

\subsection{Flaring of MAPs, flaring of thick discs, and bimodality}
Having explained how the flaring of MAPs influences the flaring of the global thick disc, and creates continuous or bimodal structures, we show in this section how these features are interconnected.

We compute a parameter that quantifies the level of flaring of all MAPs and use it to analyse the whole galaxy sample. Since what matters is not only how flared a MAP is, but also how massive it is, we
compute for each galaxy a mass-weighted average slope of MAPs as follows:

\begin{equation}
    h_{\mathrm{flaring}}=\frac{\sum_{n} m_{n} \cdot ((h_{\mathrm{outer}}-h_{\mathrm{inner}})/R_{n})}{\sum_{n} m_{n}}
\end{equation}

where $m_{n}$ is each MAP's total mass, $R_{n}$ is the radial extent of each MAP, and $(h_{\mathrm{outer}}-h_{\mathrm{inner}})$ is the difference between the MAP's scale-height at the beginning of the disc and the MAP's outermost radius. The resulting quantity that we denote as $h_{\mathrm{flaring}}$ is the averaged mass-weighted slope of all MAPs, hereafter the overall MAPs' flaring. In Fig. \ref{fig:flaring factor}, the overall MAPs' flaring is computed for all the galaxies in our sample and compared to the flaring of their respective global thick disc. For this calculation, we leave out MAPs younger than 3 Gyr and older than 9 Gyr old since these do not contribute to the flaring of the thick disc as we previously argued in Sec. \ref{sec:generalstructureMAPs}. It can be seen that, overall, MAPs always tend to flare more than their host thick discs. This is a consequence of the fact that MAPs that flare the most are not the most massive ones, especially at the flaring radii. Therefore, the global thick discs' scale-heights are going to be flattened by the MAPs that dominate the surface density at each radii, which are not the ones that flare the most. This is an example of a Yule-Simpson’s paradox, characterised as “weak” by \citealp{Minchev2019}.

Finally, we study how flaring might connected with the bimodal or continuous structure of the disc.
We represent the strength of the discs' bimodality with different marker shapes in Fig. \ref{fig:flaring factor}.
It can be seen that most of the flat thick disc galaxies have a continuous disc structure, whereas galaxies with bimodal discs tend to have larger flaring factors $h_{\mathrm{flaring}}$ and more flared thick discs.

\section{Age Gradients} 
\label{sec:agegradients}
A consequence of the combination of MAPs flaring and dominating the surface density in different disc radii is the creation of age gradients in the radial direction. In this section, we describe the age gradients in our galaxy sample and we explain how the different MAPs' flaring scenarios create radial age gradients.

\subsection{Age Gradients in the Five Cases}
\label{sec:agegradients5cases}
We start by describing how the age gradients are created in our 5 cases. For that reason, from the mid-plane to two times $h_{\mathrm{scale}}$, we create 15 equally spaced horizontal slices 0.5 kpc wide and we compute the median age of the stellar particles in the same radial bins as when computing the scale-heights. The bottom right panels of Figs. \ref{fig:flatMAPflatDISC} to \ref{fig:flaredMAPflaredDISCstack} show the median stellar ages as a function of radius, colour-coded by the constant height of the slice. A black line represents a slice 0.5 kpc wide following the thick disc's scale-height.

\begin{figure}
\centering
\includegraphics[width=\columnwidth]{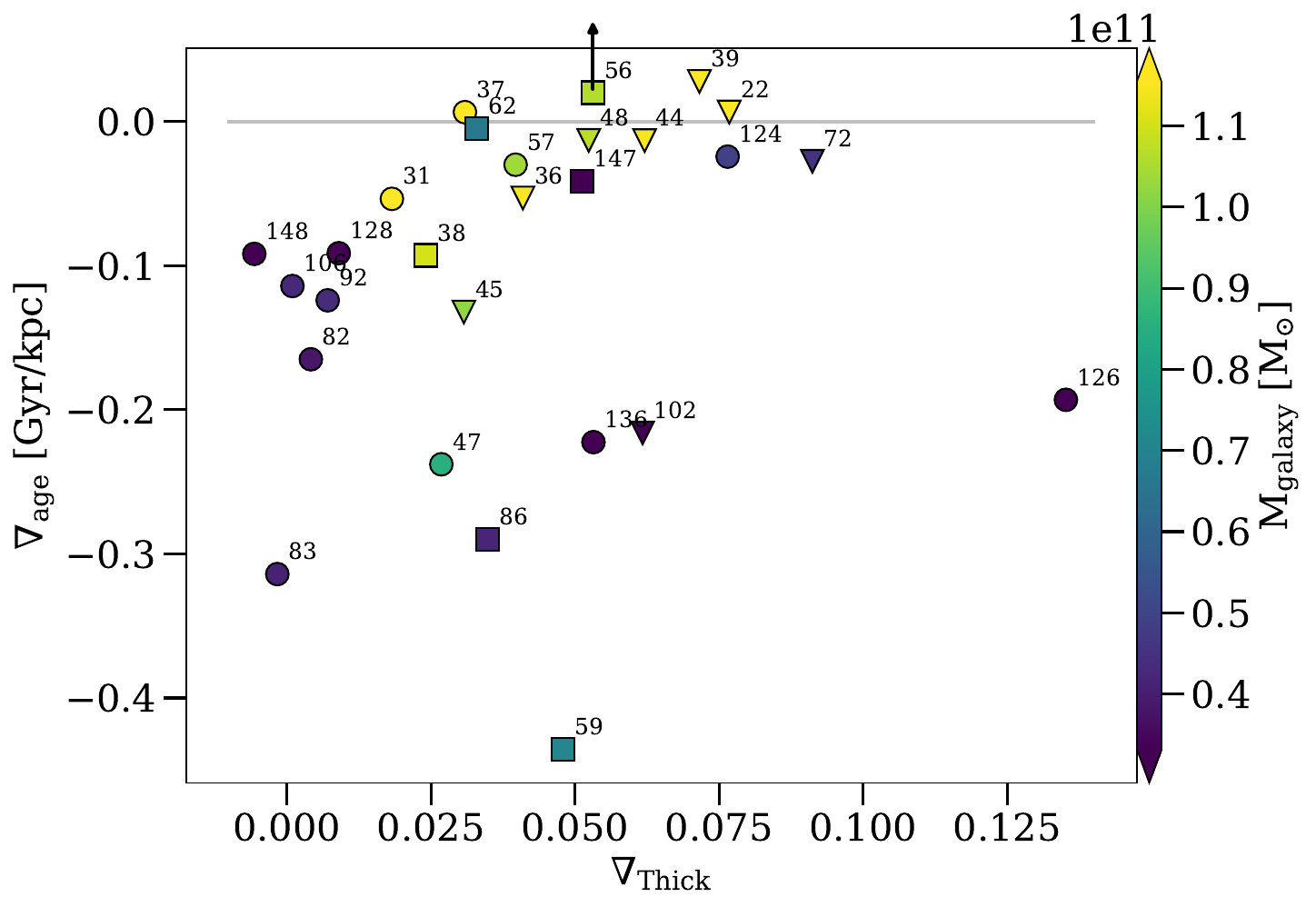}
\includegraphics[width=\columnwidth]{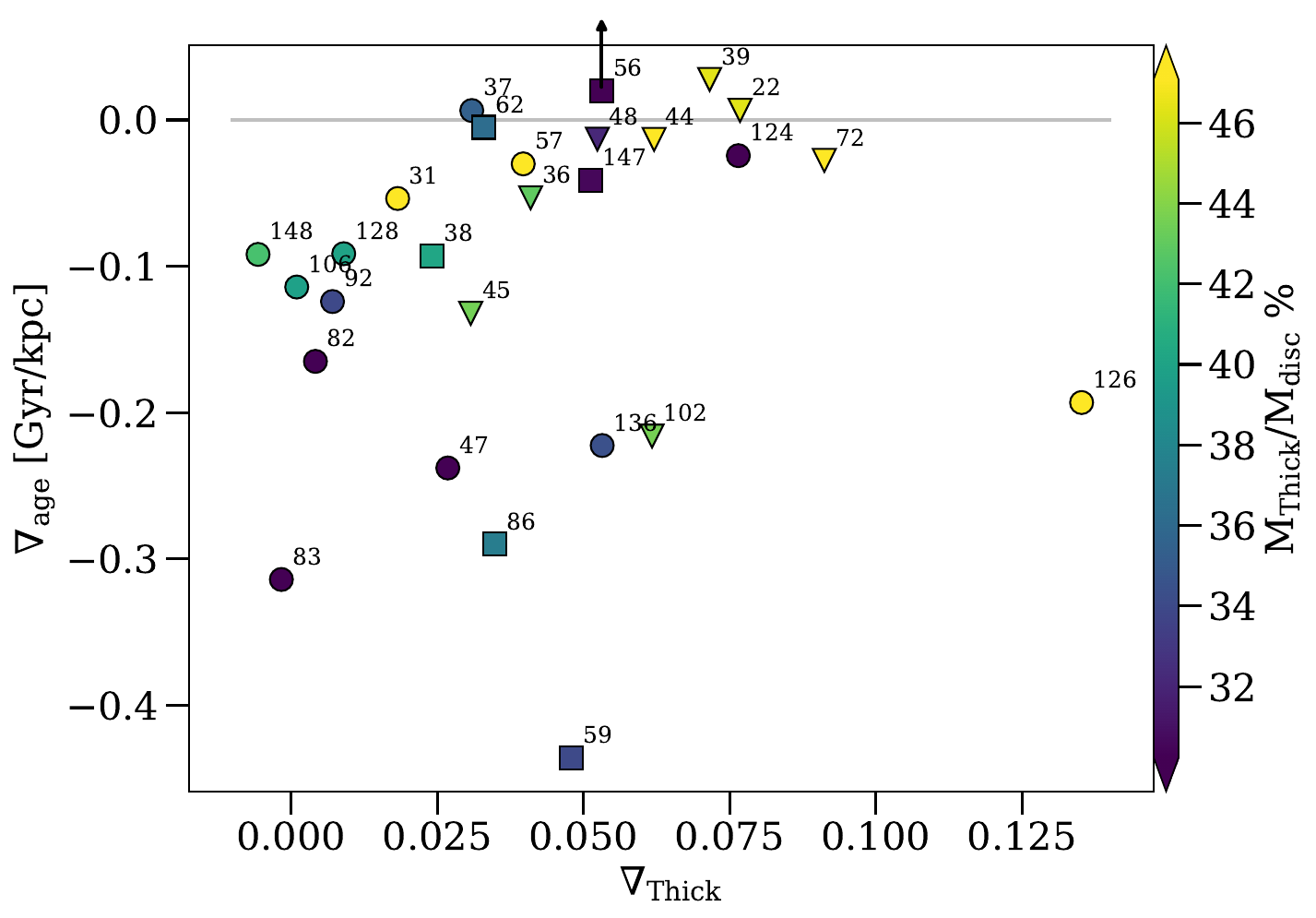}
\includegraphics[width=\columnwidth]{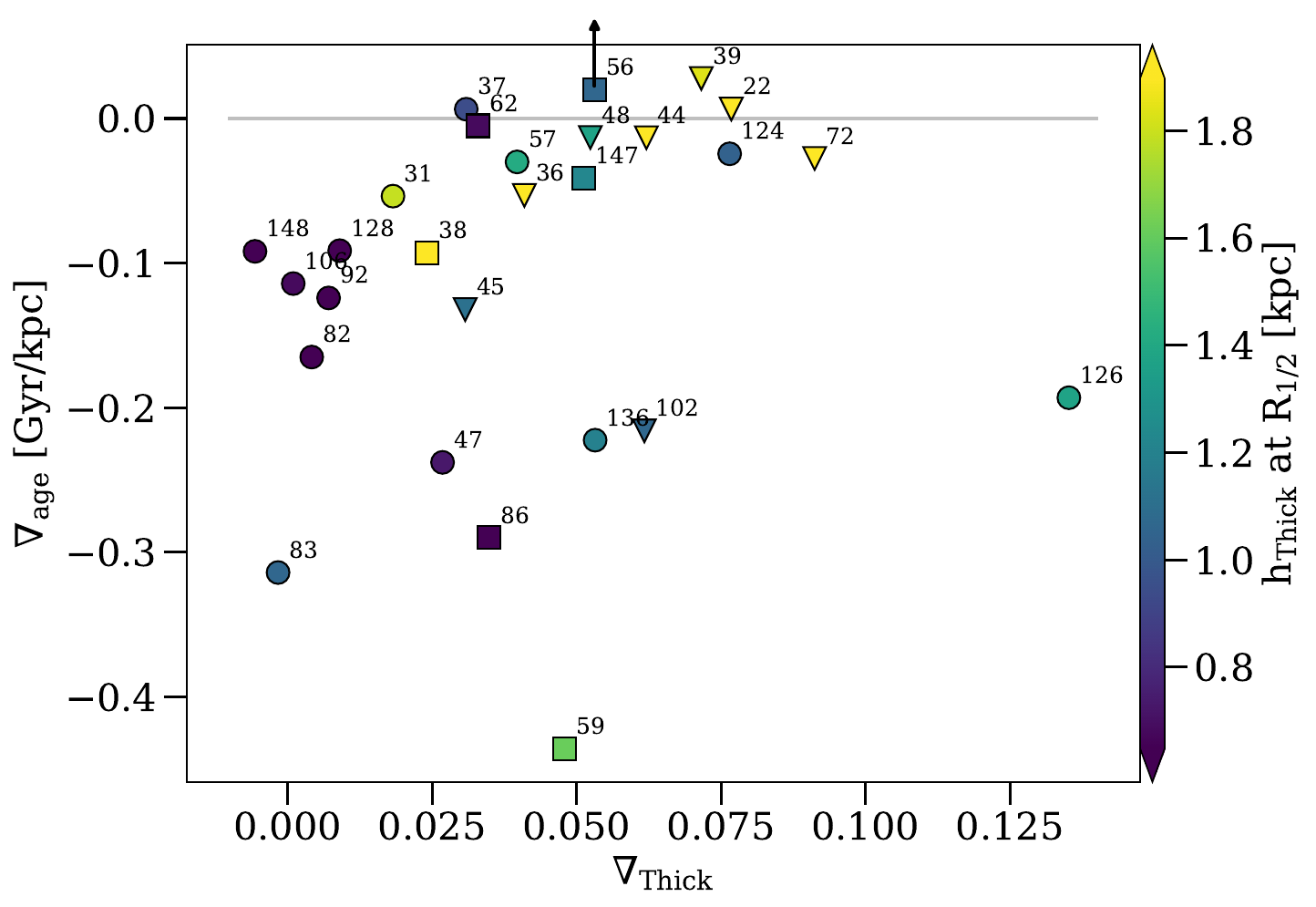}
\caption{Age gradient following the thick disc' scale-heights against thick disc gradient, colour-coded by three different variables. \textit{Top}: galaxy mass. \textit{Middle}: mass thick disc ratio. \textit{Bottom}: thickness of thick disc. The points' symbols represent the same as in Fig. \ref{fig:flaring factor}. The grey solid line highlights the constant value 0, i.e., no age gradients. All flat thick discs have an age gradient, whereas flared thick discs are more diverse, although most of them tend to have a flat age gradient.}
\centering
\label{fig:slopethickdiscslopeage}
\end{figure}

In the cases 1a and 1b (flat thick discs), the MAP dominating the surface density is changing all the time as a function of radius as shown in the upper right panel of Figs. \ref{fig:flatMAPflatDISC} and \ref{fig:flaredMAPflatDISC}. 
This makes the median stellar age decrease with radius. 
In case 1a, because both the MAPs and the thick disc are flat, the age gradient through the thick disc is very similar to the age gradient at a constant height: in the bottom right panel of Fig. \ref{fig:flatMAPflatDISC} it can be seen that the age gradient of the thick disc matches quite well the ones for heights of 0.5 and 0.75 kpc. 
Case 1b is similar to case 1a because of the small surface density of the flaring MAPs as explained previously.

Flared thick discs show a more diverse variety of age gradients. Case 2a is somewhat similar to cases 1a and 1b. 
Even though there are gaps, MAPs' scale-heights increase with age gradually (see upper left panel of Fig. \ref{fig:flaredMAPflaredDISC}). 
At the same time, younger stellar populations have higher surface density at larger radii. 
That is why at a given radius, the median stellar age increases smoothly with height, and decreases with radius. 
However, because MAPs are more extended in the vertical direction compared to cases 1a and 1b, the thick disc's scale-heights cross through only a few MAPs, which cover a small age range. 
This, together with how these MAPs' surface densities are distributed radially, creates a flat age gradient as a function of thick disc's scale-height (e.g. for galaxy g36 shown in Fig. \ref{fig:flaredMAPflaredDISC} the thick disc mean age only decreases from ~5 Gyr to ~4 Gyr over 25 kpc).
Case 2b is different because a few MAPs dominate by far the surface density over the whole disc, and that conditions the thick disc flaring as mentioned in Sec. \ref{sec:flaringmapsflaringthickdisc} (see again the upper left panel of Fig. \ref{fig:flaredMAPflaredDISCextreme}). 
In the thick disc region, the radial age gradients are mildy negative or flat above $\sim$1 kpc, where the MAPs that dominate the surface density are found. When these MAPs' densities are combined to form the thick disc, 
the age gradient following the scale-height of the thick disc is going to be quite flat too.
In the case of g39, a slightly positive age gradient results. 
Still, the gradient is small since it is determined only by a few MAPs with similar ages.

Case 2c has a different age gradient configuration from case 2a as well. 
Because several MAPs have the same scale-height, leaving a significant gap between the thin and thick disc, and because the thick disc MAPs have different radial surface density profiles, a characteristic feature is created in the age profiles (see bottom right panel of Fig. \ref{fig:flaredMAPflaredDISCstack}).
A sudden drop in age can be seen in the horizontal slices at constant height, particularly pronounced at low radius. 
This represents the transition for that horizontal slice between the MAPs on one side of the gap to the other. This drop moves out in radius as height increases, driven by the flare in the thick disc. 
This feature should not be confused with the first drop in the very inner galaxy, which is caused by the transition from the old bulge to the disc.
Also, the median age following the thick disc's scale-height decreases because the different MAPs living in the thick disc dominate the surface density at different radii as with cases 1a, 1b, and 2a. 

\subsection{Age Gradients in the Sample}
\label{sec:agegradientsWholesample}
Next, we show the age gradients of the whole sample. We compute the age gradient in the thick disc, $\nabla_{\mathrm{age}}$, by dividing the difference in age in the thick disc between $R_{\mathrm{inner}}$ and $R_{\mathrm{outer}}$ by the extent of the thick disc. 
In Fig. \ref{fig:slopethickdiscslopeage}, we represent $\mathrm{\nabla_{age}}$ as a function of the thick disc's scale-height slope, $\mathrm{\nabla_{Thick}}$, colour-coded with three different variables linked to the general properties of thick discs: mass of the galaxy in the top, thick disc mass ratio in the middle, and disc thickness in the bottom panel. The values in the colourbar indicate the 16th and 84th percentiles, so every point in dark blue or yellow belongs to the low and high tail of the distribution respectively. The marker shape represents the bimodality/continuous structure of the disc as in Fig. \ref{fig:flaring factor}.

These figures show that flat thick disc galaxies are also low mass galaxies and have thinner discs. They also tend to have low thick disc mass fractions. Finally, all flat thick disc galaxies posses an age gradient.
Above a certain level of thick disc flaring, galaxies have a wider range of behaviours, from no age gradient at all to very strong age gradients, as well as a wide range of masses, thick disk fractions and thicknesses. 
Although it might not be statistically significant, it is worth noting that about two thirds of galaxies with a flared thick disc have little or no age
gradients.
Finally, we have excluded galaxy g56 from the plot for visual purposes, as it has a gradient of 0.3 Gyr/kpc. g56 presents unusual age radial profiles that decrease with radius until close to $R_{\mathrm{outer}}$, where there is a sudden increase in age. This results in U-shaped profiles where the age at $R_{\mathrm{outer}}$ is higher than at $R_{\mathrm{inner}}$. Understanding the particular behaviour of this galaxy is not within the scope of this paper, and therefore we treat g56 as an outlier.

We find that almost all of the most massive galaxies (above 1.1 $\times$ $10^{11}$ $\mathrm{M_{\odot}}$), galaxies with the highest thick disc mass ratios (above 45\%), and the thickest discs (scale-height at $R_{1/2}$ above 1.8 kpc), have relatively flat age gradients in their thick discs.
These 3 conditions are not strictly necessary for thick discs to have flat age gradients, but we find that most galaxies which have at least one of the three characteristics have an age gradient smaller than -0.1 Gyr/kpc.

\begin{figure}
    \centering
    \includegraphics[width=\columnwidth]{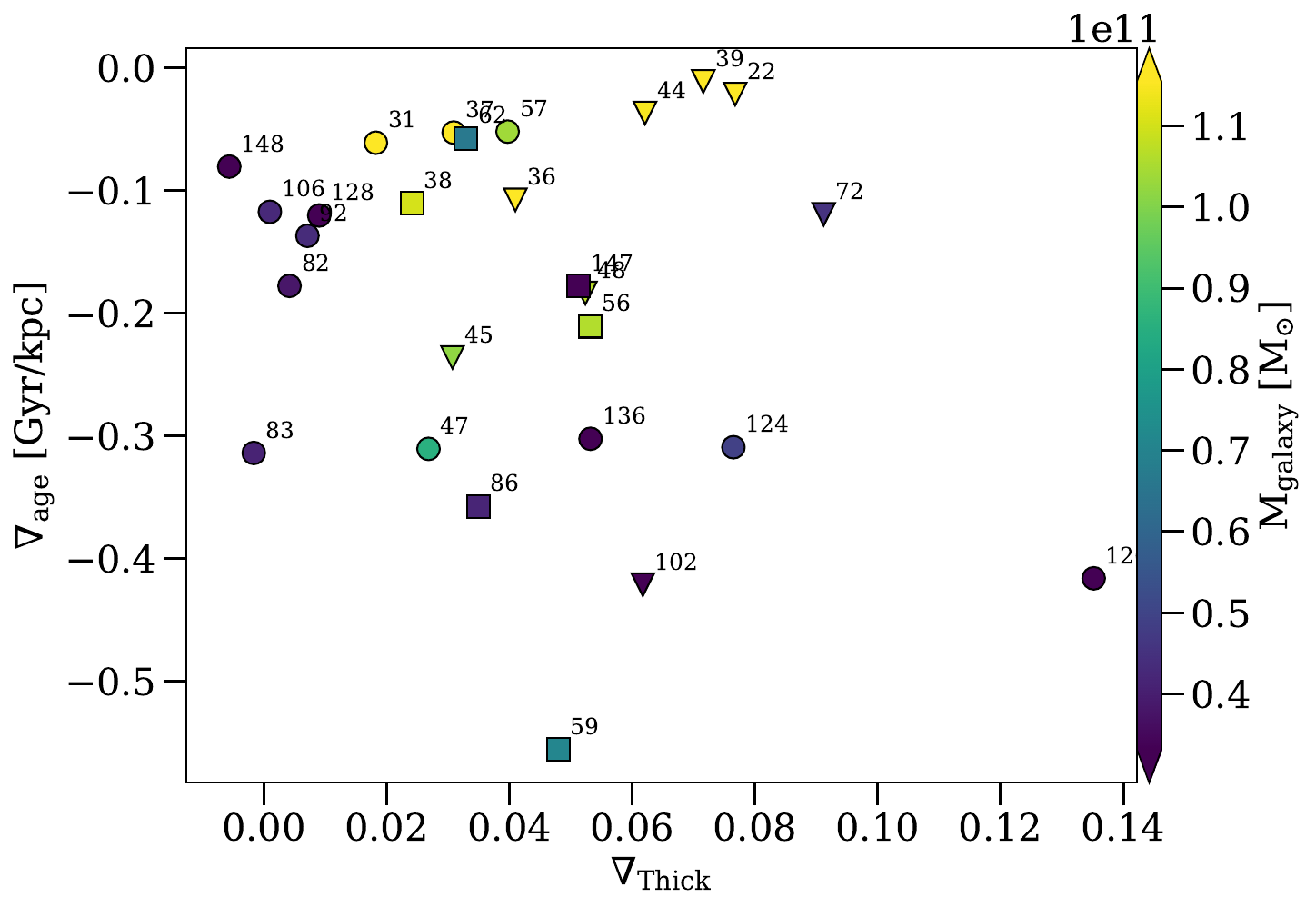}
    \caption{Age gradient at a constant height equal to the median value of $h_{\mathrm{Thick}}$ against slope of the thick disc. The markers' symbols represent the bimodality of the thick disc like in Fig. \ref{fig:flaring factor}.}
    \label{fig:constantheight}
\end{figure}

So far we have referred to radial age gradients following the thick disc's scale-heights, which imply that we follow sometimes a steep line with radius. The age gradients change of course if we take a horizontal slice in the disc at a constant height above the midplane, which is what an observer would likely do. 
We select a slice 0.5 kpc wide at a height equal to the median value of the thick disc's scale-heights, and compute the age gradient. 
As can be seen in Fig. \ref{fig:constantheight}, this does not change the age gradients measured for galaxies with a flat thick disc --- this is expected, since for those galaxies it is equivalent to measure the age at a constant height or along the thick disc scale-height. By contrast, the age gradient of galaxies with flared thick discs are generally stronger when measured at a constant height, except for the most massive galaxies with flat age gradients, which do not change much. We also note that galaxy g56 is not an outlier anymore if age gradients are measured at a constant height above the midplane.

\section{Mergers and Formation Histories}
\label{sec:mergers}

While secular evolution and radial migration can in some cases induce some disc flaring \citep{Minchev2012,Martig2014a}, interactions between discs and satellite galaxies are one of the main mechanisms causing flaring. How much flaring is induced depends on multiple factors such as the orbit and mass of the satellite, and the bulge fraction of the main galaxy
\citep{Kazantzidis2008ColdAccretion,Kazantzidis2009ColdAccretion,Qu2011CharacteristicsLengths,Moetazedian2016ImpactDisc}.
It is not within the scope of this work to explore meticulously how those different factors produce different effects on disc dynamics and flaring, but we can observe some of their most noticeable effects, and explain some configurations of thick disc, MAPs, and age gradients.

\begin{figure}
    \centering
    \includegraphics[width=\columnwidth]{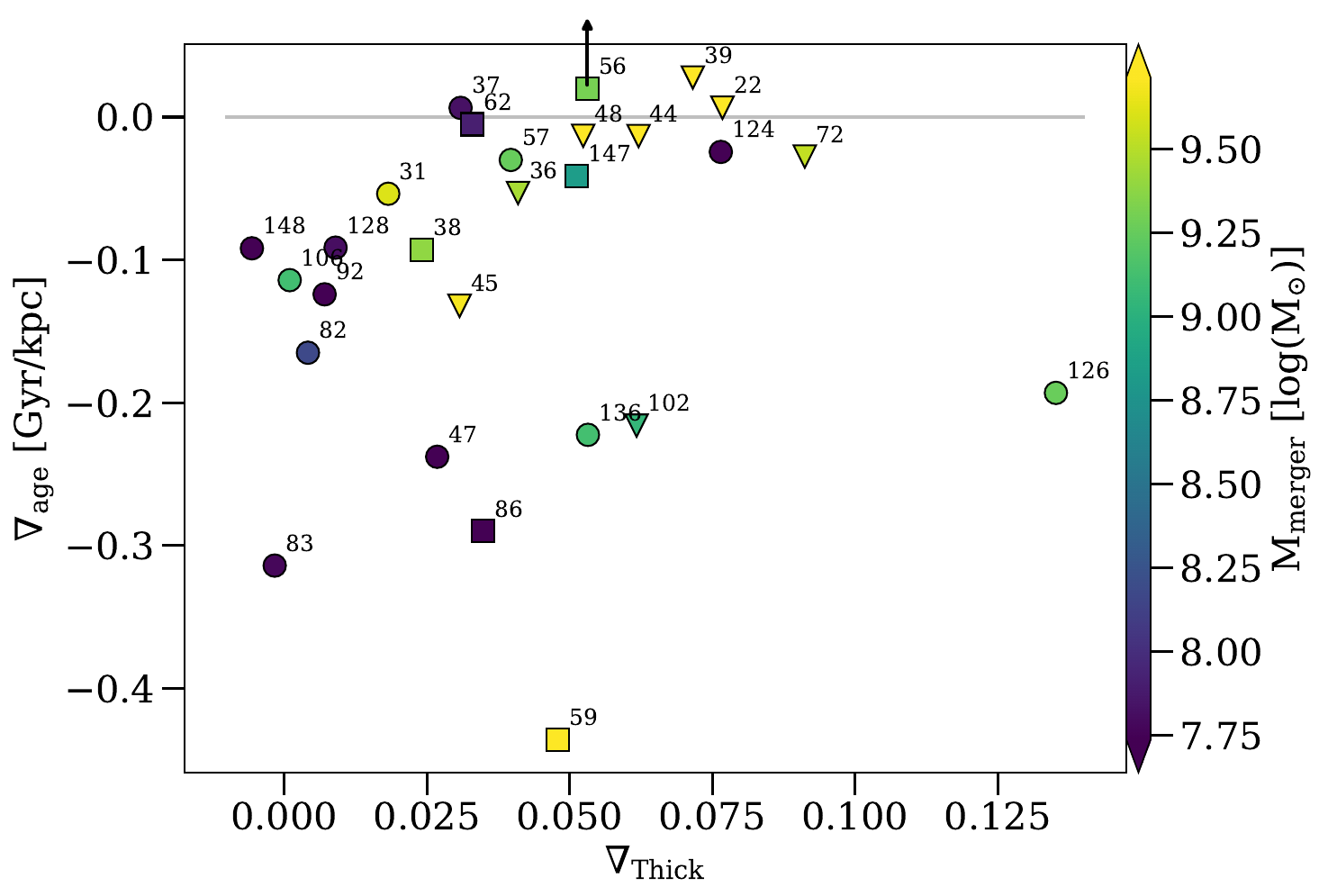}
    \includegraphics[width=\columnwidth]{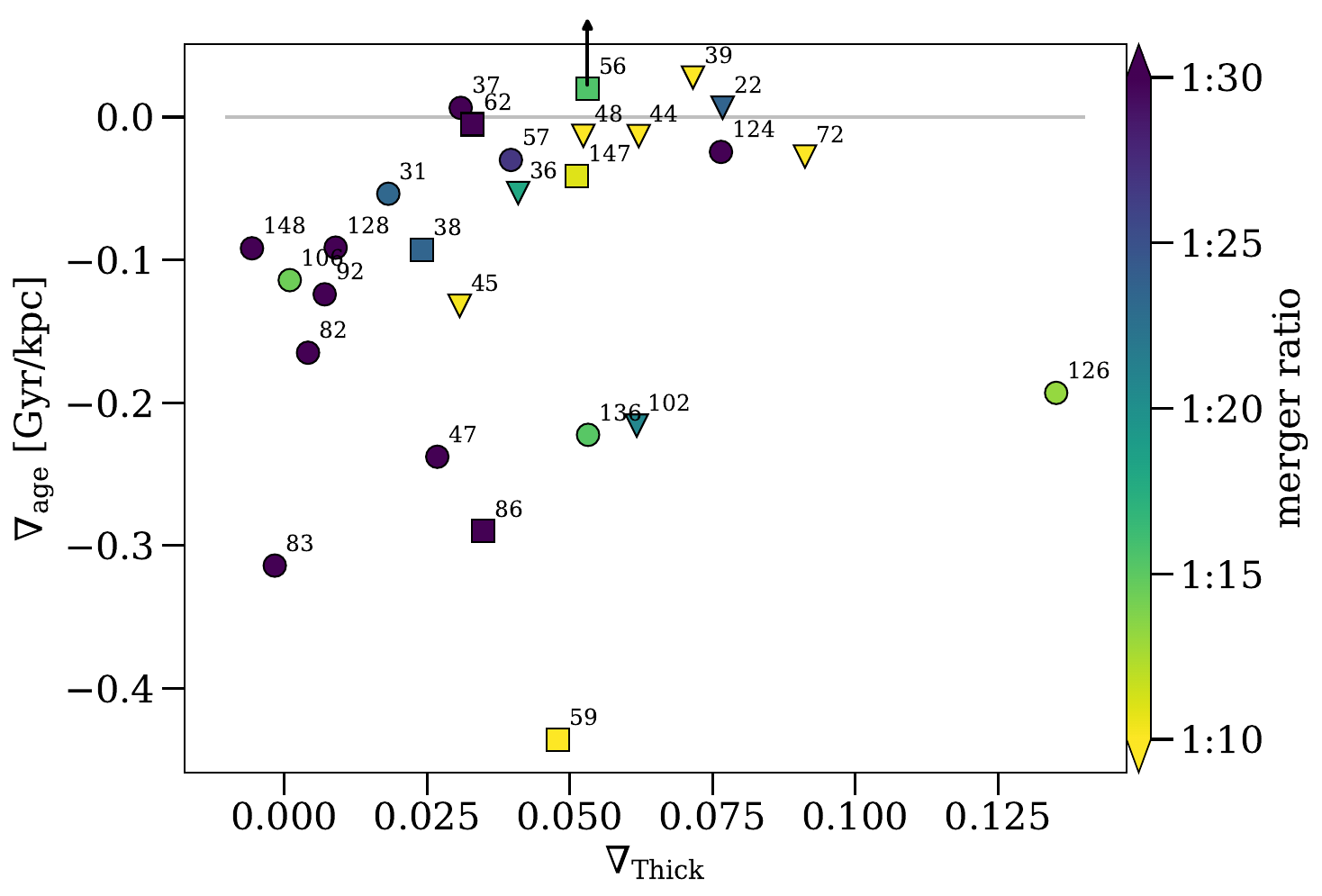}
    \caption{Age gradient following the thick disc's scale-height against the slope of the thick disc. \textit{Top}: colour-coded by the mass of the most massive merger in the last 9 Gyr. \textit{Bottom}: colour-coded by the mass ratio of the most massive merger in the last 9 Gyr (defined as the ratio between the stellar mass of the satellite and the main galaxy at the time of the merger) The symbols represent the bimodality of the thick disc like in Fig. \ref{fig:flaring factor}. Galaxies with stronger mergers tend to have more flared thick discs with flatter age gradients.}
    \label{fig:scatterplotsmergers}
\end{figure}

The effect mergers have on thick disc flaring and age gradients are varied. For galaxies like g36 in case 2a, represented in Fig. \ref{fig:flaredMAPflaredDISC}, a merger will flare all MAPs born up to that point, whereas MAPs born after are going to be considerably less flared, creating a gap \citep{Martig2014}. This is not necessarily enough to produce a flat age gradient along the thick disc. In galaxies like g102 in case 2c, the merger will make all the MAPs born up to that point flare with the same scale-heights, creating a thin/thick disc bi-modality. Yet, there will be still an age gradient along the thick disc since the MAPs populating it still dominate the surface density at different radial distances. In other cases however, mergers will make one or few MAPs overmassive compared to the rest, creating a flat age gradient in the thick disc like galaxy g39 in case 2c.

In Fig. \ref{fig:scatterplotsmergers} we show the age gradient along the thick disc $\mathrm{\nabla_{age}}$ as a function of the thick disc's scale-height slope $\mathrm{\nabla_{Thick}}$ as we did in Fig. \ref{fig:slopethickdiscslopeage}, but this time colour-coded by the stellar mass of the most massive merger in the last 9 Gyr in the upper panel, and the mass ratio of that merger in the bottom panel. 
The shape of the points has the same meaning as in Figs. \ref{fig:flaring factor} and \ref{fig:slopethickdiscslopeage}. 
The stellar mass of the satellites is computed right before their first interaction with the disc of the main galaxy. Thus, we are not considering the initial mass of satellites and their disruption as they cross the galactic halo, which is a process that could be potentially altered by numerical resolution \citep{VanDenBosch2018DisruptionFiction}. We also refer the reader to \citealp{Martig2012,Martig2014a} for a more detailed discussion of some of the resolution tests that we have performed.

We find that galaxies with flat thick discs all had very quiescent merger histories: all of those  galaxies (except g106 that had a 1:15 merger around 8 Gyr ago) only have mergers smaller than 1:30 in the last 9 Gyr.
Radial age gradients along the disc are a natural consequence of an inside-out disc growing formation scenario. 
As the disc forms and grows gradually, younger MAPs are born and dominate the density towards the outer regions. 
For flat thick disc galaxies, which generally have quiescent merger histories, this configuration will remain intact as the MAPs' flaring is either minimal or happens at the radii with low surface density. 
Therefore, flat thick disc will tend to have a radial age gradient along the disc, which is what we observe in Figs. \ref{fig:slopethickdiscslopeage} and \ref{fig:scatterplotsmergers}. 

Flared thick disc galaxies show more diversity. Galaxies with more massive mergers tend to have flat age gradients (less than --0.1 Gyr/kpc) and also tend to have some degree of flaring (more than 0.02).
All the galaxies having merged with satellites more massive than 4.5 $\times$ $10^9$ $\mathrm{M_{\odot}}$ show either mild or strong bimodality, although there is no strong correlation between the degree of bimodality and the mass of the satellite. For instance, g59 underwent a massive merger but only shows mild bimodality. 
This could be due to different factors, like the merger's orbit: exploring this is not within the scope of this paper, and future work will be needed on this matter.
The bottom panel of Fig. \ref{fig:scatterplotsmergers} shows similar results. Most galaxies that underwent mergers with a mass ratio higher than 1:10 seem to have more flared thick discs with flat age gradients. These galaxies also show strong thin/thick disc bimodality. 
By contrast, we also find galaxies with a quiescent merger history but a flared thick disc and a flat age gradient. 
For these galaxies, there must be other mechanisms besides mergers causing both the flaring of the thick disc and the flat age gradient.
In some of these galaxies, MAPs are born already with some degree of flaring, which has been reported in other cosmological simulations (e.g. \citealp{Navarro2018}).
As for the others, MAPs were born in flatter configuration.
The source of flaring for these MAPs can be internal heating mechanisms or external agents such flybys or multiple minor mergers throughout the life time of the galaxy. 
However, most galaxies in our sample have no more than one merger with a mass ratio above 1:30, and the few with more mergers (up to 4 above a merger mass ratio above 1:30) do not seem to be different from the rest of the galaxy sample. 

Even though mergers are not the only mechanism for disc flaring, they play a key role in our sample. 
We do observe that if the merger is massive enough, or the mass ratio between the merger and the main galaxy is high enough, they can produce a flared thick disc, a flat age gradient, and a bimodal thin/thick disc structure. 
On the other hand, we observe that galaxies with a flat thick disc  have a relatively quiescent merger history.

\section{Conclusions}

In this work we explored the structure of galactic thick discs using a sample of 27 simulated galaxies in their cosmological context \citep{Martig2012}.
\citealp{Minchev2015ONDISKS} explained how Mono-Age Populations (MAPs) create the thin and thick disc by a combination of three factors: the MAPs' level of flaring, the individual MAPs' surface density, and how these two change with radius. This interplay creates age radial gradients in the thick disc (as observed in the Milky Way, \citealp{Martig2016AWAY}). 
We expand on the work of \citealp{Minchev2015ONDISKS}, and use a sample of 27 simulated galaxies to explore: i) how the flaring of MAPs and the flaring of the global thick disc  are connected, ii) whether the thin and thick discs are distinct components or a continuum, iii) what kinds of age gradients flaring MAPs form, iv) and how merger histories can potentially influence all the above. 
Our conclusions from this study can be summarised as follows:

\begin{itemize}
    \item MAPs older than around 9 Gyr are very concentrated in the centre of the galaxy and barely contribute to the flaring of the thick disc. They do, however, increase the scale-height of the thick disc quite uniformly at all radii. MAPs younger than around 3 Gyr  are very concentrated in the galactic plane and barely flare, therefore they will not contribute to the flaring of the thick disc either. Flaring of the thick disc is mainly driven by stellar populations between 3 and 9 Gyr old.

    \item Flat thick discs form when MAPs barely flare or when, due to inside-out formation, the flared MAPs do not carry a lot of surface density at the flaring radii compared to younger, less flared MAPs. 
    The MAPs' scale-heights increase smoothly as a function of age, creating a continuous thin/thick disc structure as was found for the Milky Way \citep{Bovy2012}, although mono-age and mono-abundance populations can be quite different \citep{Minchev2016THEDISK}. We find that all flat thick discs exhibit a radial age gradient like the Milky Way \citep{Martig2016AWAY}, and they are galaxies with low mass (e.g. $\mathrm{M_{galaxy}} \leq 5\times 10^{10} \mathrm{M_{\odot}}$), thinner global discs (e.g. average $h_{\mathrm{Thick}} \leq 1 \mathrm{kpc}$), and lower thick disc mass ratios (e.g. $\mathrm{\mathrm{M_{thick}/\mathrm{M_{disc}}}} \leq 40\%$). 
    They also tend to have more quiescent merger histories, with low mass or low mass ratio mergers, or mergers only at early times. In our sample, 12 galaxies have flat thick discs, which corresponds to 44\% percent of the sample.

    \item Flared thick discs form when MAPs carry a significant amount of surface density where they flare. 
    The flaring of the global thick disc can be driven by a sequence of different MAPs at different radii, or only MAPs spanning a couple of Gyrs if those MAPs dominate the surface density throughout the disc, or they all share the same scale-heights. 
    If one of the two last cases happens, then a bimodal structure is created and thin and thick disc are distinct components in terms of the stellar populations inhabiting them. 
    This effect is directly related with the merger history as \citealp{Martig2014a} pointed out using the same galaxy sample.
    Flared thick discs are more diverse than their flat counterparts in terms of age radial profiles, although a high fraction of them show a small or flat age radial profile following their thick disc's scale-heights. 
    This is especially the case for galaxies with high mass, high thick disc mass fractions, and the thickest global discs. 
    We also find that galaxies that underwent very massive mergers or with high mass merger ratio tend to have flat age radial profiles. 
    In our sample, 15 galaxies have flared thick discs, which corresponds to 56\% percent of the sample.
\end{itemize}

Our sample of simulated galaxies thus shows a great diversity of thick disc structures, even though we only studied a relatively limited halo mass range and restricted ourselves to galaxies in isolated environments. We find a group of 6 galaxies that show characteristics close to the Milky Way, like radial age gradients in their thick discs, and minimal flaring of the global thick discs' scale-heights. These galaxies belong to the lower end of the distribution in our sample in terms of stellar mass ($\mathrm{M_{galaxy}} \leq 5\times 10^{10} \mathrm{M_{\odot}}$), have average thick disc scale-heights ($h_{\mathrm{Thick}} \leq 1 \mathrm{kpc}$), and are in the lower half of the distribution in terms of thick disc stellar mass ratio ($\mathrm{\mathrm{M_{thick}/\mathrm{M_{disc}}}} \leq 40\%$).
The rest of the sample, i.e. flared thick disc galaxies, is a lot more diverse, depending on the galaxies' formation histories. Our results cannot be directly compared to galaxies in the Fornax cluster \citep{Pinna2019TheAccretion,Pinna2019TheEnvironment}, but they could be used to interpret, for instance, the discovery in NGC 7572 of a very massive and flared thick disc \citep{Kasparova2020An7572}: this probably results from a massive merger. Our simulations suggest that thick discs can be successfully used to probe galaxies' merger histories, which we will study in more details in future papers.
As more and more data becomes available for thick discs in nearby galaxies, we will soon be able to compare the Milky Way and its neighbours,  to interpret their different structures, and to connect those with their formation histories.

\section*{Acknowledgements}

The authors would like to thank the referee for very useful comments, which greatly helped improve this manuscript.
We would also like to thank Teresa Antoja for useful discussions, and Simon Pfeifer and Thomas Sedgwick for their continuous support from the beginning of this project.
We would like to acknowledge a LIV.DAT doctoral studentship supported by the STFC under contract[ST/P006752/1].
The LIV.DAT Centre for Doctoral Training (CDT) is hosted by the University of Liverpool
and Liverpool John Moores University / Astrophysics Research Institute.

\section*{Data Availability}
The data underlying this article will be shared on reasonable request to the corresponding author.

\bibliographystyle{mnras}
\bibliography{paper} 

\bsp	
\label{lastpage}
\end{document}